\documentclass[prd,aps]{revtex4}

\usepackage{epsf}

\begin{document}
\title{ Late-time dynamics of rapidly rotating black holes }
\author{Kostas Glampedakis\footnote{email: spxcg@astro.cf.ac.uk}}
\affiliation{ Department of Physics and Astronomy, Cardiff University
 P.O. Box 913, Cardiff, CF2 3YB, UK}

\author{Nils Andersson\footnote{email: N.Andersson@maths.soton.ac.uk}}

\affiliation{ Department of Mathematics, University of Southampton, 
Southampton SO17 1BJ, UK}
\date{\today}

\begin{abstract}
We study the late-time behaviour of a dynamically perturbed rapidly 
rotating black hole. Considering an extreme Kerr black hole, we show that 
the large number of virtually undamped quasinormal modes (that exist for 
nonzero values of the azimuthal eigenvalue $m$) combine in such a way that 
the field (as observed at infinity) oscillates with an amplitude that
decays as $1/t$ at late times. This is in clear contrast with the standard 
late time power-law fall-off familiar from studies of non-rotating 
black holes. This long-lived oscillating ``tail'' will, however, not 
be present for non-extreme (presumably more astrophysically
relevant) black holes, for which we find that many quasinormal modes 
(individually excited to a
very small amplitude) combine to give rise to an exponentially 
decaying field. This result could have implications for the detection 
of gravitational-wave signals from rapidly spinning black holes, since the 
required theoretical templates need to be constructed from linear combinations of many modes. Our main results are obtained analytically, but we support the conclusions with numerical time-evolutions of the Teukolsky equation. 
These time-evolutions provide an interesting insight into the notion that 
the quasinormal modes can be viewed as waves trapped in the spacetime region
outside the horizon. They also suggest that a  plausible mechanism for the 
behaviour we observe for extreme black holes is the presence
of a ``superradiance resonance cavity'' immediately outside the black hole.
\end{abstract}

\maketitle


\section{Introduction}

\subsection{A brief background}

The generic response of a black hole to 
dynamical perturbations was first investigated more than thirty years 
ago. Vishveshwara \cite{vishu} was the first 
to observe, in simple numerical experiments where a Schwarzschild black hole
was pelted by Gaussian wave-pulses, the ``ringing'' 
associated with the exponentially damped quasinormal modes (QNMs). 
Shortly thereafter, 
the late time power-law fall-off (that all perturbative fields 
decay as $t^{-2l-3}$ in the Schwarzschild geometry) was discovered 
by Price \cite{price}. Since these first studies, a considerable 
body of work has established the importance of these two phenomena 
for black hole physics (see \cite{novikov} for a review).
 
We now know that the late-time tail is due to 
radiation backscattered by the  weak 
gravitational potential in the far-zone\cite{GPP}. 
It is independent of the central object's  
strong-field features (such as the presence of an event horizon), and will 
consequently be the same for (say) a neutron star and a black hole with the
same gravitational mass. 
In contrast, the QNMs represent 
radiation scattered in the strong-field regime.
Hence, they provide the means for making a clear distinction between
a neutron star and a black hole. A neutron star has many families of 
fluid pulsations modes, an observation of which would provide 
useful constraints on the supranuclear equation of state \cite{astero}.
In addition, there is a set of modes that arise because gravitational
waves can be temporarily trapped by the spacetime curvature caused by the
presence of the star. These modes share some qualitative features 
with the QNMs of a black hole, which  
arise as waves are temporarily trapped in the vicinity of the 
black hole's gravitational potential peak (located near the 
unstable photon orbit at $r=3M$ in the Schwarzschild case). The modes 
are analogous to scattering resonances in quantum physics and thus
decay exponentially in time. In a detailed study, 
Leaver \cite{leaver1} identified the QNMs as complex-frequency poles
 of the relevant Green's function (see also \cite{na}). 
In addition, he showed that the power-law tail, that dominates the 
black holes dynamical response once the QNMs have died away, originates
from the presence of a branch cut in the Green's function
(customarily placed along the negative imaginary
frequency axis).

Most previous studies of QNMs and power-law tails have  focused on
non-rotating black holes, and our understanding of this case has reached 
a mature level. The same can not be said about the rotating case, however. 
A few years ago, the QNMs had been calculated also for Kerr black holes \cite{leaver2,onozawa}, but there were  no actual calculations demonstrating the presence of power-law tails. 
Several recent developments have improved 
our understanding of dynamical rotating black holes. Of particular relevance 
has been an effort to develop a reliable framework for perturbative
time-evolutions of Kerr black holes \cite{tcode1,tcode2,tcode3,tcode4}. 
There has also been recent efforts to analytically approximate 
the late-time power-law tails for Kerr black holes \cite{hod,barack1,barack2}. 
The main conclusions of these studies support the standard picture: The QNM 
signal (typically dominated by the slowest damped mode) gives way at 
late times to a power-law fall-off. Although the QNM frequencies and 
the late-time power-law are slightly
altered by rotation, the results are qualitatively similar to the 
Schwarzschild ones. 
For example, for a massless  scalar field 
Ori and Barack~\cite{barack1,barack2} 
 predict a fall-off (for non-extreme Kerr black holes)
proportional to 
$ t^{-l-|m| -3 -q}$ where $q=0/1$ for even/odd $l+m$. 
This should be compared to the Schwarzschild
result  $t^{-2l-3}$. 
The difference can be  attributed 
to the rotation-induced coupling of the various $l$-multipoles
in the signal.


\subsection{Motivation: The observability of QNM signals}

With a new generation of gravitational-wave detectors due to reach their
projected sensitivities within the next few years, 
the question whether we can realistically hope to do 
``black-hole spectroscopy'' by detecting QNM signals following (say) the 
formation of a black hole in a supernova is of obvious interest. 
In principle, 
such a detection should enable us to infer the black-hole mass and angular
momemtum and therefore provide an unambiguous identification of 
astrophysical black holes. 

For slowly rotating black holes this presents a serious challenge. 
To demonstrate this in a simple way we consider the gravitational wave signal 
associated with a certain QNM. 
Far away from the black hole  the associated 
 flux follows from the standard formula
\begin{equation}
F = { 1\over 16\pi} \vert \dot{h} \vert^2 = {1\over 4\pi r^2 }
{dE \over dt} .
\end{equation}
Using this together with
\begin{equation}
{dE \over dt} = {E \over 2\tau_e}
\end{equation}
where $\tau_e$ is the e-folding time of the QNM, and assuming a
monochromatic wave (of frequency $f$) such that $\dot{h} = 2\pi f
h$ (although we note that this assumption is not justified 
for rapidly damped QNMs), we get
\begin{equation}
h =  \left( {E \over \tau_e} \right)^{1/2} {1\over \pi rf}
\label{flux}\end{equation} 
Finally, we estimate the effective
amplitude achievable after matched filtering as
\begin{equation}
h_{\rm eff} = h\sqrt{N} = h\sqrt{f\tau_e} 
\label{heff1}
\end{equation}
where $N$ is the number of detected cycles of the signal.
Parameterising this result we arrive at 
 (cf. similar estimates for pulsating stars \cite{astero})
\begin{equation}
h_{\rm eff} 
\approx 4.2\times 10^{-24} \left( {\delta\over 10^{-6}}
 \right)^{1/2} \left(  { M \over M_\odot } \right)
\left( {15{\rm Mpc} \over r} \right) 
\end{equation} 
where $\delta$ is the radiated energy as a fraction of the black hole's 
mass $M$. For a Schwarzschild black hole
the frequency of the radiation depends on the 
black-hole mass as $f \approx 12  ({M_\odot/ M}) {\rm kHz }$. 
Given these relations, and recalling the estimated sensitivity of the 
generation of ground based 
detectors that is under construction (LIGO, VIRGO, GEO600 and TAMA), 
we see that 
it is going to be difficult to see QNM signals from slowly rotating solar-mass black holes. The mode-signals are basically too rapidly damped. But there are reasons 
not to despair. First of all,  the situation is much more
favourable for low-frequency signals from supramassive black holes
 in galactic nuclei and detection with LISA, 
the space-based interferometric gravitational-wave antenna. 
Secondly, there have been recent indications that  ``middle weight''
 black holes, with masses in the range $100-1000M_\odot$,  
exist \cite{mediumbh1,mediumbh2}. For such black holes 
the most important QNMs would radiate at frequencies where the new generation 
of ground based detectors reach their peak sensitivity ($\sim 100$~Hz). 
If there are such  black holes in our galaxy, and they are
dynamically perturbed by some external agent, we may 
hope to detect the resultant QNM ringing in the future. 

It  has been suggested that QNM signals 
from rapidly rotating black holes 
would be easier to detect than what the above estimate suggests. 
This belief is based on the fact that some QNMs become very 
long lived as $a\to M$ (where $0\le a \le M$ is the rotation 
parameter of the black hole). QNM calculations predict the 
existence of an infinite set of essentially undamped modes in the 
extreme Kerr limit \cite{leaver2,onozawa,detweiler}, and previous
investigations into the detectability 
of QNM signals have focused on the slow damping of these modes 
\cite{echeverria,finn,flanagan}. 
It is easy to see how the decreased damping of the modes 
may increase the detectability considerably.
But it is also easy to understand that a
 slowly damped mode is only easier to detect than a short-lived one if 
the modes are excited to a comparable amplitude. This then
forces us to address (rather difficult) issues regarding QNM excitation
\cite{na,nollert}.
In particular, we must investigate whether it is easier to 
excite a slowly damped QNM than a short-lived one. Intuitively, one 
might expect this not to be the case. In similar physical 
situations the build-up of energy in a long-lived resonant 
mode takes place on a time-scale similar to the eventual mode
damping. Consequently, it ought to be quite difficult to excite a QNM that 
has characteristic damping several times longer than the dynamical 
timescale of the excitation process. 
This argument suggests that the amplitude of each 
long-lived mode ought to vanish in the limit $a\to M$ when the e-folding 
time of the mode increases dramatically. Indications that 
this is the case were provided by Ferrari and Mashhoon some years ago 
\cite{ferrari}.

To illustrate this further, we connect (\ref{heff1}) to the 
amplitude of the QNM under
consideration. We do this by  assuming that the mode-signal can be
represented by
\begin{equation}
h \approx {A  e^{- t/\tau_e} \over r} \sin(2\pi f t) \ .
\end{equation}
Then we can use (\ref{flux}) to infer the ``amplitude'' $A$ that
corresponds to a certain ``total energy radiated through the
mode'' $E$. We get
\begin{equation}
E = {\pi^2  f^2 \tau_e  \over 4 }  A^2 \ ,
\end{equation}
and the effective gravitational-wave amplitude
\begin{equation}
h_{\rm eff} = {\sqrt{f \tau_e} \over 2} A \approx \left( 1 - {a\over
M } \right)^{-3/10} {A \over \sqrt{2\pi} } \ .
\label{heff}\end{equation}
Here we have used an empirical approximation for the frequency 
and damping rate of the
fundamental $l=m=2$ mode \cite{echeverria}. When the 
effective amplitude is expressed in this form  
it is quite clear that a decrease 
in $A$ can easily compensate for the increase in $\tau_e$ that
 occurs as the black hole spins up. In other words, in order 
to correctly discuss the detectability of the Kerr QNMs one 
must necessarily investigate this balance. This provides the  
prime  motivation for the present work.


\subsection{A brief summary of previous work}

In a previous paper \cite{prl} we discussed some 
issues concerning the long-lived QNMs of a Kerr black hole. Our 
results essentially followed by studying a suitably defined 
excitation coefficient (defined in terms of the asymptotic 
behaviour of the perturbed field, see Section~IIE of the present
paper),  $A^{\rm out}/\alpha_n$.
Provided that one can calculate this quantity 
(which is far from trivial \cite{na}) for a given QNM as the 
black hole spin varies one can estimate the associated change in level of 
excitation of the mode. This follows since
 we can express the effective amplitude (\ref{heff})
associated with a single QNM (with complex frequency $\omega_n$) as
\begin{equation}
h_{\rm eff} =  \frac{\sqrt{f\tau_e}}{2}A \sim  \sqrt{ { \mbox{Re } \omega_n  
\over \mbox{Im }
\omega_n } } {A^{\rm out}\over \alpha_n} \ .
\label{heff2}
\end{equation}

Using results derived in the main part of this paper
we find, for an extreme Kerr black hole,  that
\begin{equation}
h_{\rm eff} \sim e^{-n\pi/2\delta}
\end{equation}
where $\delta$ is a positive real constant (see Section IID for its 
definition),
and the least damped modes correspond to large values of the integer $n$. 
From this we immediately deduce that the ``effective detectability'' is 
exponentially small for these modes.

Similar evidence is provided by recently obtained
numerical data \cite{modepaper} relevant for the full range of $a$.
Before we discuss these results we
recall that the QNMs of a non-rotating black hole are located symmetrically 
with respect to the imaginary frequency axis in the complex $\omega$-plane. 
For non-zero $a$ this symmetry is broken and co- and counter-rotating modes
will be affected in different ways. The counter-rotating modes remain rapidly
damped, while the co-rotating modes become very slowly 
damped as $a\to M$. 
These conclusions follow from eg. Figure~3 in \cite{leaver2}. 

In Figure ~\ref{fig1} we compare the ``effective gravitational-wave amplitude''
obtained from (\ref{heff2}) for the 
slowest damped co- and counter-rotating QNMs of a Kerr black 
hole perturbed by a massless scalar field.  From the 
figure we see that, even though the co-rotating mode
is much longer lived, its ``detectability'' decreases
significantly as $a\to M$. In fact, our results 
suggest that the effective amplitude of the slowest damped QNM for a 
rapidly rotating black hole may be as much as three orders of magnitude 
smaller than the amplitude 
of the corresponding mode of a Schwarzschild black hole. This is
 clearly very bad news as far as detecting the various QNMs 
of a rapidly spinning black hole is concerned.

\begin{figure}[tbh]
\caption{ An assessment of the ``detectability'' of  Kerr 
black hole QNMs as $a\to M$.
The effective amplitude $h_{\rm eff}$ (as estimated from Eq. (\ref{heff2}))
is plotted as a function of the spin
parameter $a$. We compare the slowest damped co-rotating 
QNM to  the slowest damped counter-rotating one.  
The figure shows that the co-rotating mode (which becomes 
very long lived in the near extreme case) has a much smaller excitation 
coefficient than the counter-rotating mode as $a\to M$.}
\label{fig1}
\end{figure}

However, one further complication must be accounted for.
Our argument shows that each individual QNM  has a small amplitude 
in the limit of a rapidly spinning black hole. But we need to account 
for the fact that a 
large number of modes approach the same limiting frequency as $a\to M$. 
Conceivably, these modes
could interfere constructively and result in a considerable signal.
Indeed, our previous estimates \cite{prl} 
suggest that, despite the fact that the individual 
long-lived modes are excited to a negligible level their collective 
contribution to the late-time signal is significant. 
This then leads to a  late-time behaviour of a rapidly 
rotating black-hole that is not at all well described in the 
standard terms (as a signal mainly consisting of
the leading QNM and the late-time power-law tail). 
Instead, for an extreme black hole the long-lived modes 
completely dominate the late-time behaviour, by giving rise to  
an oscillating signal with an amplitude that decays as $1/t$.
In this paper we present the complete 
derivation of this result, 
and  discuss the extent to which it remains relevant
also for astrophysical (non-extreme) black holes. 


\subsection{Organisation of this paper}

The remainder of the paper is organised as follows.
 Part II contains (Section IIA \& IIB) the formulation of the 
initial-value problem in the Kerr geometry, a brief description of the
analytic properties of the relevant Green's function in the complex 
frequency plane (Section IIC), and a discussion of the long-lived 
Kerr QNMs (Section IID). Part II then concludes with the calculation 
of the late-time field due to these modes, which leads to  the main 
result of this paper (Section IIE). 
Section~III concerns numerical time-evolutions and  QNM signal reconstructions
that amplify and strengthen the analytical results. 
In Part IV we discuss a physical interpretation of the results, 
and introduce the concept of a ``superradiance resonance cavity''.
Our conclusions are in Part V. Some technical details are
discussed in Appendices~A-C. 
In Appendix A we present approximate solutions of the Teukolsky equation
for near extreme black holes first obtained by Teukolsky and 
Press \cite{teuk2}.
In Appendix B the analytical properties of the solutions of the Teukolsky equation are examined in some detail. Finally, in Appendix C, the realistic 
case of gravitational perturbations is briefly discussed. Throughout the paper 
geometrised units $G=c=1$ have been adopted.


\section{The Cauchy Problem in Kerr Geometry}

In the last decade several results have emphasized the surprising 
accuracy of black-hole perturbation theory in situations 
where one would intuitively expect it to fail. The most celebrated
example of this is the close-limit approximation  for black hole
collisions devised by Pullin and Price \cite{climit}. It has become clear
that perturbative methods provide not only a benchmark test   
for fully nonlinear numerical relativity, but that it often 
offers a reliable alternative or complement \cite{lousto}.
In fact, in many problems black-hole perturbation theory captures
 a significant part of the physics, although it obviously will not 
provide any insights into the non-linear behaviour (unless 
taken to higher orders \cite{campa}).

Perturbative studies are attractive because of their 
relative transparency. Once we make the assumption that the perturbative 
field is weak in the sense that its contribution to the 
spacetime curvature can be neglected,  the black hole evolution equations 
can  be cast in the form of a wave equation with a 
complicated effective potential \cite{novikov,chandra}. This makes 
the initial-value problem amenable to an analytical (or 
comparatively simple numerical) 
treatment. Furthermore, in studies where the emphasis is on the 
qualitative behaviour one can make further simplifications by 
considering a massless scalar field. Since the more realistic
cases of  gravitational or electromagnetic fields are governed by 
perturbation equations very similar to the scalar wave equation, one would 
expect the simple toy problem (recall that no scalar fields have not yet 
been observed in Nature) to provide generally relevant information.
The advantage of the scalar field problem is simply that one does not have 
to go through the several steps involved in recreating (say) the 
perturbed metric from the employed variables. Consequently, 
we mainly focus on the dynamics of a massless
scalar field in the Kerr background in this paper. But since some
of our conclusions may prove relevant for future gravitational-wave 
detection we discuss the gravitational field problem in  
Appendix C. The main conclusions drawn from this Appendix is that all the 
results from the scalar field problem remain relevant also 
in the gravitational case.


\subsection{A formal solution}

The standard formalism for studying Kerr black hole perturbations is 
based on the Newman-Penrose approach. As was first shown by Teukolsky 
almost thirty years ago \cite{teuk1}, the evolution of any perturbative 
field can be described by a single ``master equation'', which in the case
of a scalar field takes the form
\begin{eqnarray}
& & \left[ \frac{(r^2 + a^2)^2}{\Delta} -a^2 \sin^{2}\theta \right]
\frac{\partial^2 \Phi}{\partial t^2} + \frac{4Ma r}{\Delta}
\frac{\partial^2 \Phi}{\partial t \partial \varphi}  \nonumber \\
& & + \left[ \frac{a^2}{\Delta} -\frac{1}{\sin^2\theta} \right]
\frac{\partial ^2 \Phi}{\partial \varphi^2} -\frac{\partial}{\partial r}
\left( \Delta \frac{\partial \Phi}{\partial r} \right) - 
\frac{1}{\sin\theta} \frac{\partial}{\partial \theta}
\left( \sin\theta\frac{\partial \Phi}{\partial \theta} \right) = 0 
\label{mastereq}
\end{eqnarray}
where we have used  standard Boyer-Lindquist coordinates.
In (\ref{mastereq}) we have also used $\Delta= r^2 -2Mr +a^2$. 
The two horizons of the black hole then follow from 
$\Delta =0$ and can readily be found to be $r_{\pm}= M \pm \sqrt{M^2 -a^2}$. 
In order to simplify (\ref{mastereq}) we introduce 
the tortoise coordinate $r_{\ast}$, which is defined as
\begin{equation}
\frac{dr_{\ast}}{dr}= \frac{r^2 +a^2}{\Delta}
\end{equation}
Integrating, we have for $a<M$
\begin{equation}
r_\ast = r + { 2Mr_+ \over r_+ - r_- } \ln \left( {r\over r_+} -1
\right) - { 2Mr_- \over r_+ - r_- } \ln \left( {r\over r_-} -1
\right) \ .
\end{equation}
and for $a=M$
\begin{equation}
r_\ast= r -\frac{2M^2}{r-M} + 2M\ln \left( {r\over M} -1 \right)
\end{equation}
Furthermore, the axisymmetry of the Kerr background allows us to 
immediately separate out the dependence on the azimuthal angle $\varphi$
by introducing
\begin{equation}
\Phi= (r^2 + a^2)^{-1/2}\sum_{m=-\infty}^{\infty} \Phi_{m}
(t,r,\theta) e^{im\varphi}
\end{equation}
The function $\Phi_{m}(t,r,\theta) $ then satisfies
\begin{eqnarray}
&&\left[ { (r^2+a^2)^2 \over\Delta} - a^2\sin^2\theta \right] \frac{\partial
^2 \Phi_{m} }{\partial t^2} +\left[ \frac{4iMamr}{\Delta}\right]  
\frac{\partial \Phi_{m}}
{\partial t} - { (r^2 + a^2)^2 \over \Delta} {\partial^2 \Phi_{m}  \over \partial
r_\ast^2} \nonumber \\ 
&&+  { (r^2 + a^2)^2 \over \Delta} \left[ {dG \over dr_\ast}
+ G^2 \right] \Phi_{m} -m^2\left[ \frac{a^2}{\Delta}
-\frac{1}{\sin^2\theta}\right]
\Phi_{m} -\frac{1}{\sin\theta} \frac{\partial}{\partial \theta}
\left(\sin\theta \frac{\partial \Phi_{m} }{\partial
\theta}\right)=0 
\end{eqnarray}
where $G=r\Delta/(r^2+a^2)^2$.

In order to proceed further we
use a Laplace-type integral transform; 
\begin{eqnarray}
\hat{F}(\omega)&=& \int_{0^-}^{+\infty} F(t) e^{i\omega t} dt \ , \\
F(t) &=&  {1\over 2\pi} \int_{-\infty+ic}^{+\infty+ic} \hat{F}(\omega)
e^{-i\omega t} d\omega \ ,
\end{eqnarray}
where $c$ is some (positive) constant, to bring the problem into
the frequency domain.
If the initial data for the field $\Phi_{m}(t,r,\theta)$ is given 
by
\begin{eqnarray}
\Phi_0 &=& \Phi_{m}(0,r,\theta) \ , \\
\dot{\Phi}_0 &=& \left. {\partial \Phi_{m}(t,r,\theta) \over \partial t}
\right|_{t=0} \ ,
\end{eqnarray}
we then get the transformed equation
\begin{equation}
{\cal D} \hat{\Phi}_{m} +{\cal L} \hat{\Phi}_{m} = {\cal S} (\Phi_0,
\dot{\Phi}_0) \ . \label{unsep}\end{equation}
Here we have defined the two differential operators
\begin{equation}
{\cal D} = {(r^2+a^2)^2 \over \Delta} \left\{ {\partial^2 \over
\partial r_\ast^2} + \left[ {K^2 + (2am\omega -a^2 \omega^2)
\Delta \over (r^2+a^2)^2}
-{dG \over dr_\ast} - G^2 \right]\right\} \ ,
\end{equation}
where $K=(r^2 + a^2)\omega - am$,  and
\begin{equation}
{\cal L} = {1 \over \sin \theta} {\partial \over \partial
\theta}\left(\sin \theta {\partial \over \partial \theta} \right)
 + a^2 \omega^2 \cos^2\theta - {m^2 \over \sin^2 \theta} \ ,
\end{equation}
and the source term is
\begin{eqnarray}
{\cal S} (\Phi_0, \dot{\Phi}_0) =
 \left [ \frac{i\omega  (r^2+a^2)^2 -4iamMr}{\Delta}
- i\omega a^2 \sin^2\theta  \right ] \Phi_0 
- \left[{(r^2+a^2)^2 \over \Delta} - a^2\sin^2 \theta \right] \dot{\Phi}_0 
\end{eqnarray}

Finally, we would like to separate the radial and angular dependencies.
To do this we assume that we can find a function $S(\theta)$ such that
\begin{equation}
{\cal L} S + E S = 0 \ ,
\label{angeq}\end{equation}
where $E$ is an eigenvalue that ensures that $S$ is regular at
both $\theta=0$ and $\theta =\pi$. The angular
eigenfunctions are then defined via a Sturm-Liouville problem for $E$.
The theory of such problems tells us that the eigenfunctions form
a complete, orthogonal set on the interval $0\le \theta \le \pi$
for each combination of $a\omega$ and $m$. Since the functions
must limit to the standard spin-weighted spherical harmonics 
(once the factor $e^{im\varphi}$ is included) as $a\omega \to 0$ it 
makes sense to label each 
eigenfunction by the integer $l$ (with $l\in [0,\infty]$).
In other words, we denote the solutions to (\ref{angeq}) as
$S_{l m}(\omega,\theta)$.  
We further require
that the eigenfunctions are normalised in such a way that
\begin{equation}
\int_0^\pi \sin \theta S_{l m} \bar{S}_{nm} d\theta = \delta_{l n} \ ,
\end{equation}
with the bar denoting complex conjugation.
By assuming that  the left- and right-hand sides  of (\ref{unsep})
are both expanded in the complete set of angular functions, i.e. that
$\hat{\Phi}_{m} = \sum_{l=|m|}^\infty R_{l m}(\omega, r) 
S_{l m}(\theta, a\omega) $ etcetera, we now get

\begin{equation}
{d^2 R_{l m} \over dr_\ast^2} + \left[ {K^2  + (2am\omega
-a^2\omega^2 - E)\Delta \over (r^2+a^2)^2} -{dG \over dr_\ast} -
G^2 \right] R_{l m} = {\Delta \over (r^2 + a^2)^2} {\cal
S}_{l m}(\omega, r) \ , 
\label{radeq}
\end{equation} 
where
\begin{equation}
{\cal S}_{l m}(\omega, r) = \int_0^\pi \bar{S}_{l m}(\theta,a\omega) 
{\cal S} (\Phi_0, \dot{\Phi}_0) \sin \theta d\theta \ .
\end{equation}
We note that the effective radial ``potential'' is
\begin{equation}
Q(r,\omega)= {K^2  + (2am\omega -a^2\omega^2 - E)\Delta \over (r^2+a^2)^2} 
-{dG \over dr_\ast} -G^2
\label{potent}
\end{equation} 

A solution to (\ref{radeq}) can be obtained via the  Green's
function technique. Specifically, we  find that
\begin{equation}
R_{l m}(\omega, r) = \int_{-\infty}^{\infty} {\Delta(r^\prime)
G(r_\ast^\prime, r_\ast) \over [(r^\prime)^2 + a^2]^2} {\cal
S}_{l m}(\omega, r^\prime) dr_\ast^\prime \ ,
\end{equation}
where the primed variables represent the source point and the
unprimed represent the location of the observer. The 
required Green's function can be
expressed in terms of two linearly independent solutions to the
homogeneous version of (\ref{radeq}). These solutions are defined
by their behaviour close to the event horizon and at spatial
infinity in such a way that
\begin{equation}
u^{\rm up} \sim \left\{ \begin{array}{ll} B^{\rm out} e^{ikr_\ast} 
+ B^{\rm in} e^{-ikr_\ast}
\quad \mbox{as } r\to r_+ \ , \\ e^{i\omega r_\ast} \quad \quad 
\mbox{as } r\to +\infty \ , \end{array} \right.
\label{up}
\end{equation}
and
\begin{equation}
u^{\rm in} \sim \left\{ \begin{array}{ll}
e^{-ikr_\ast} \quad \mbox{as } r\to r_+ \ , \\ A^{\rm
out} e^{i\omega r_\ast} + A^{\rm in} e^{-i\omega r_\ast} \quad
\mbox{as } r\to +\infty \ .
\end{array} \right.
\label{in}
\end{equation}
Here
\begin{equation}
k = \omega - {ma \over 2Mr_+} = \omega - m \omega_+ \ ,
\end{equation}
where $\omega_+$ is the angular velocity of the event horizon.
Finally,  we have
\begin{equation}
G(r_\ast^\prime, r_\ast) = - {1 \over 2i\omega A^{\rm in}} \left\{
\begin{array}{ll} u^{\rm up}(r_\ast) u^{\rm in}(r^{\prime}_{\ast}) \ , 
\quad r^\prime_\ast < r_\ast \ , \\ u^{\rm in}(r_\ast) u^{\rm
up}(r^\prime_\ast) \ , \quad r^{\prime}_{\ast} > r_\ast \ .
\end{array} \right.
\end{equation}

and the scalar field follows
after inverting the integral transform:
\begin{equation}
\Phi_{m}(t,r,\theta) =  {1\over 2\pi} \sum_{l=|m|}^\infty
\int_{-\infty+ic}^{+\infty+ic}
 R_{l m}(\omega,r) S_{l m}(\theta,a\omega)
e^{-i\omega t} d\omega \ .
\label{mfield}
\end{equation} 
In this way we have arrived at a formal solution to the initial
value problem for the Teukolsky equation.


\subsection{The asymptotic approximation}

Our aim  now is to analyse the Kerr problem further and assess
the level of excitation of the various QNMs. It is generally
accepted that this is a far from trivial problem, see \cite{na,nollert}
for a detailed discussion, and we will not try to
provide a complete solution here. Instead, we will study a 
particular model scenario that allows us to make further
simplifications. Thus we  use the so-called 
``asymptotic approximation'' that was first introduced in \cite{na}, 
i.e. assume that i) the observer is situated far away from
the black hole, and ii) the initial data has support 
only in the far zone (but 
inside the observer). The first of these assumptions corresponds to
$r_\ast>>M$, while the second means that $M<<r^\prime_\ast<<r_\ast$.
Given these assumptions we can use the asymptotic behaviour (\ref{up})
and
(\ref{in}) of the solutions to the Teukolsky equation to construct the 
required Green's function analytically. 

The asymptotic approximation is convenient, but is it
realistic to introduce these two assumptions? 
The assumption that the 
observer is located at large distances from the black hole 
should be relevant for most problems of astrophysical interest
(and we  note that one could readily devise an analogous approximation
for the case when the observer is located close to the horizon).
Furthermore, there is certainly a class of problems for which the 
second assumption holds. Namely, when well-defined wave pulses
fall onto the black hole. This is is a standard scenario considered
in qualitative studies of black-hole dynamics, but it is 
not a scenario likely to give rise to  detectable astrophysical
gravitational waves. For the astrophysical scenarios
that would seem the most relevant, gravitational collapse to form a
black hole and the coalescence of two black holes following 
binary inspiral, the assumption that the ``initial data'' is 
located far from the horizon will not hold. But again, one could devise a 
similar approximation in which the data has support only close to the horizon.
Of course, we know from previous studies of related problems \cite{na}
that the main QNM excitation arises from data in the region near the peak 
of the perturbative potential (which is located near the unstable 
photon orbit). For such data the asymptotic approximation will not be 
accurate, and it is therefore likely to be 
of limited use in realistic scenarios.

Anyway, once we introduce the asymptotic approximation, we get
\begin{equation}
G(r^\prime_\ast, r_\ast) \approx - {e^{i\omega r_\ast} \over 2i\omega}
\left\{ {A^{\rm out} \over  A^{\rm in} } e^{i\omega r^\prime_\ast} + 
 e^{-i\omega r^\prime_\ast} \right\} \ .
\label{Gasymp}\end{equation}
Then, the solution to the radial Teukolsky equation follows from
\begin{equation}
R_{l m}(\omega, r) \approx  - {e^{i\omega r_\ast} \over 2i\omega}
\int_{-\infty}^{\infty} {\Delta(r^\prime) \over [(r^\prime)^2 + a^2]^2}
\left\{ { A^{\rm out} \over  A^{\rm in} } e^{i\omega
r^\prime_\ast} + e^{-i\omega r^\prime_\ast} \right\} {\cal
S}_{l m}(\omega, r^\prime) dr^\prime_\ast \ .
\end{equation}
and expression (\ref{mfield}) for the field becomes
\begin{equation}
\Phi_{m}(r,\theta,t) \approx -\frac{1}{2\pi} \sum_{l=m}^{\infty} 
\int_{-\infty+ic}^{+\infty+ic} \frac{e^{-i\omega(t-r_\ast)}}{2i\omega}
 S_{l m}(\theta,a\omega) \left [ \frac{A^{\rm out}}{A^{\rm in}} 
{\cal I}_{l m}^{+}(\omega)
+ {\cal I}_{l m}^{-}(\omega) \right ] d\omega
\label{mfield2}
\end{equation}
Here we have defined the two integrals
\begin{equation}
{\cal I}_{l m}^{\pm}(\omega) = \int_{-\infty}^{+\infty} 
\frac{ \Delta(r^\prime)}
{[(r^\prime)^2 + a^2 ]^2} e^{\pm i\omega r^{\prime}_{\ast}}
{\cal S}_{l m}(\omega, r^\prime) dr^\prime_\ast  
\end{equation}
Given specific initial data, the calculation of the field requires evaluation 
of the frequency integral in (\ref{mfield2}).
If we make further simplifications by assuming 
static initial data, i.e. $\dot{\Phi}_0=0$, we arrive at
\begin{equation}
{\cal I}_{l m}^{\pm}(\omega) \approx i\omega \int_{0}^{\pi} 
d\theta \sin\theta \bar{S}_{l m}(\theta,\omega) 
\int_{-\infty}^{+\infty} e^{\pm i\omega r_\ast^{\prime} } 
\Phi_{0}(r^\prime,\theta) dr_\ast^{\prime}
\end{equation}


\subsection{Working in the complex frequency plane}

In practice, the frequency integrals in (\ref{mfield}) and (\ref{mfield2}) 
cannot be directly evaluated analytically. It is a straightforward 
task to evaluate them numerically for any acceptable initial data, 
but this does not provide much insight into the qualitative features of the
solution. As is by now well known, a useful alternative is to analyze  
the problem in the complex $\omega$-plane \cite{leaver1,na}.
To do this, we bend the  integration contour into the lower half
of the complex frequency plane and use Cauchy's theorem to 
calculate the integral as a sum over the contributions from 
the deformed contour plus a sum of the residues of the poles 
encircled by the contour. Naturally, we need to understand the  
analytical properties of the Green's function in order to 
follow this approach.

The properties of the Schwarzschild black hole Green's function 
have been exhaustively investigated, but the Kerr case has not 
attracted nearly as much attention \cite{hw}. We find that we must  
 consider two different cases depending on whether we 
are interested in an extreme ($a=M$) or a non-extreme ($a<M$) black hole. 
The analytical properties of the Green's function for the latter case are
exactly as in the Schwarzschild problem: An infinite set of 
simple poles, corresponding to the QNMs, are located in the lower half 
plane and a branch point (originating from the $u^{\rm up}$ function) 
at $\omega=0$ leads to a branch cut that is usually placed along the 
negative imaginary axis, cf. Figure~1  of \cite{na}. 
The existence of this branch cut is 
related to the asymptotic behaviour of the effective potential 
in the Teukolsky equation. Namely the fact that
 the potential (beyond the centrifugal term) falls off as $ \ln(r_\ast)/r_{\ast}^{3}$
as $r_\ast \rightarrow +\infty$, 
see \cite{ching} for a detailed discussion of this point.

For $a=M$ we find an additional feature: The function $u^{\rm in}$ 
now has a branch point at $\omega= m\omega_{+}$. This means that 
there will in principle be a 
second branch cut in the complex $\omega$-plane.
This time, it is the behaviour of the potential at 
$r_\ast \rightarrow -\infty$ 
that leads to the presence of a branch cut. The effective 
potential near the horizon falls off as $ 1/r_\ast $ instead of 
exponentially (as in the non-extreme case). 
The presence of this second branch cut is discussed further in Appendix B.


\section{The long-lived Kerr quasinormal modes}

In analogy with standard scattering theory, the quasinormal modes
can be interpreted as ``resonances'' of the black hole's gravitational 
field. This is apparent since they correspond to the poles
of the Green's function which, as can be seen from 
(\ref{Gasymp}), means that they correspond to
\begin{equation}
\left. \frac{A^{\rm in}}{A^{\rm out}} \right|_{\omega=\omega_{n}}= 0 \ .
\label{qnmdef}
\end{equation}
The physical interpretation of this  is that the QNMs correspond to waves that are purely outgoing at spatial infinity, while at the same time corresponding 
to purely ingoing waves crossing the event horizon. Given this definition, 
it is straightforward to see that ``conservation laws'' force the solutions to (\ref{qnmdef}) to be complex-valued \cite{ferrari} (in fact, for a Kerr black hole, a QNM could exist on the real axis as well, but all attempts to find such modes have failed \cite{teuk2,hw}). 
It further turns out that there is an infinite set of such mode solutions for each  given combination $(l,m)$, corresponding to complex eigenfrequencies $\omega_n = \omega_n(M,a,l,m)$. The QNM spectra of non-rotating black holes have been exhaustively discussed in the literature and are by now well 
understood \cite{novikov}. In contrast, there have only been a few studies
of the corresponding Kerr problem. Kerr black hole QNMs were first calculated by Detweiler \cite{detweiler}. The most detailed calculations were 
done by Leaver \cite{leaver2} a few years later, and more recently Onozawa \cite{onozawa} extended Leaver's continued fraction approach to discuss
near-extreme black holes. Kerr QNMs and excitation coefficients have also been
calculated by the present authors by integrating suitably chosen 
phase-functions along paths in the  
complex $r_{\ast}$-plane \cite{modepaper}.  
These studies lead to the following conclusions: As the black hole spins up, 
each Schwarzshild QNM (belonging to an infinite family of modes 
for each $l$) splits into $2l +1$ separate modes (due to the breakdown of 
spherical symmetry which makes the various $m$ solutions distinct). How the modes are influenced by the black hole's rotation essentially depends on whether they are co- or counter-rotating with the black hole. One finds that modes that co-rotate ($m>0$) with the black hole become longer lived as $a$ increases (i.e. the imaginary part of the  QNM frequency 
is significantly reduced). At the same time the oscillation frequency increases. This effect is most pronounced for the $l=m$ modes. In fact, as 
$a\rightarrow M$ these modes all approach the $\omega= m\omega_{+}$ point on 
the real frequency axis, cf. Figure~3 in \cite{leaver2}. This behaviour 
can be understood if we note that the $l=m$ modes can be interpreted as 
being ``associated with'' 
the black hole's equatorial plane (cf. the symmetry of 
the spherical harmonics). If such a mode was co-rotating with the
black hole it would experience maximal frame-dragging effects 
(see \cite{ferrari} for a particularly transparent treatment in the 
eikonal approximation). If we recall that the QNMs can be 
associated with waves being trapped in the vicinity of the 
unstable photon orbit we see that, as the black hole approaches 
extreme rotation rates, the co-rotating modes will be located 
closer and closer to the horizon until at $a=M$ when the frequency of these
``equatorial'' modes is a multiple of the rotation frequency 
of the horizon ($\omega_+$). This then makes the extremely slow damping of 
these QNMs natural. Furthermore, it is easy to understand why other QNMs are less effected by rotation. Other co-rotating modes, with $l \neq m$, will 
be dragged along with the black hole but they never become extremely long-lived (essentially because they are not symmetric with respect to the equatorial plane). And the   ``counter-rotating'' modes are much less affected by 
rotation, remaining close to their Schwarzschild counterparts as the black 
hole spins up. These qualitative results are the same 
for all perturbative fields (gravitational, electromagnetic or scalar).


\subsection{Approximating the QNM frequencies}

We aim to investigate the extent to which the extremely long-lived
QNMs of a rapidly spinning black hole are excited by a generic
perturbation. As outlined in the Introduction, the main motivation for this
study is the question whether one should expect these modes to be present
in the gravitational-wave signal from astrophysical black holes.
If this were the case, such signals would be considerably easier
to detect than the short bursts that would be characteristic of  a 
Schwarzschild black hole. 

As demonstrated by Detweiler \cite{detweiler}, the long-lived QNMs 
can be obtained analytically using an approximation to the 
Teukolsky equation that is valid for $a\approx M$ and 
$\omega \approx m\omega_{+}$. This approximation was first introduced by  Teukolsky and Press \cite{teuk2}, and for completeness we reproduce their calculation in Appendix A. Adopting their notation, we define new variables
\begin{eqnarray}
x &=& {r - r_+ \over r_+ } \ , \label{ptvar1} \\ 
\sigma &=& {r_+ - r_- \over r_+} \ , \\ 
\tau &=& M \left( \omega - m\omega_{+} \right) \ , \\ 
\hat{\omega} &=& \omega r_+ \, \\
\delta^2 &=& 4\hat{\omega}^2 -1/4 -\lambda \ ,
\label{PTvars}
\end{eqnarray}
where $\lambda= E + a^2 \omega^2 -2ma\omega $.
This means that the extreme Kerr limit corresponds to $\sigma \to 0$.
Furthermore, one can show that $\delta$ is almost purely real (imaginary) for
$l=m$ $(l \neq m)$. As can be seen from the equations below, this fact 
distinguishes the $l=m$ case and leads, from a computational
point of view, to the presence of the long-lived QNMs.

From the equations in Appendix~A we see that
the condition (\ref{qnmdef}) corresponds to the following 
equation;
\begin{equation}
- { \Gamma(2i\delta) \Gamma(1+2i\delta)  \over \Gamma(-2i\delta)
\Gamma(1-2i\delta)}\left[{\Gamma(1/2-2i\hat{\omega}-i\delta) \over
\Gamma(1/2-2i\hat{\omega}+i\delta) } \right]^2 =(-2i\hat{\omega}\sigma)^{2i\delta} 
{\Gamma(1/2+2i\hat{\omega}+i\delta-4i\tau/\sigma) \over
\Gamma(1/2+2i\hat{\omega}-i\delta-4i\tau/\sigma) } \ .
\label{modecond}
\end{equation}
The left-hand side of this equation has a well-defined limit as
$a\to M$ and $\omega \to m\omega_{+}$. We represent that limit by
\begin{equation}
\mbox{LHS } = qe^{i\chi} \ .
\end{equation}
Meanwhile, we see that we cannot have a consistent solution unless
$\tau/\sigma\to \infty$ as $a\to M$. Then  the
right-hand side of the mode-condition can be written
(using Stirling's formula)
\begin{equation}
\mbox{ RHS } = (-8\hat{\omega}\tau)^{2i\delta} \ .
\end{equation}
In other words, a QNM must be a solution to
\begin{equation}
f(\omega) = (-8\hat{\omega}\tau)^{2i\delta} - qe^{i\chi} = 0 \ .
\end{equation}
Using $-8\hat{\omega}\tau = \rho e^{i\zeta}$ we see that
solutions follow from (remembering that $\delta$ is real for $l=m$
and $\omega \to \omega_+$)
\begin{equation}
\rho = \exp\left[ { \chi - 2n\pi \over 2\delta} \right]
\end{equation}
\begin{equation}
\zeta = - {1\over 2\delta} ln q \ .
\end{equation}
From this, we can conclude that there are an infinite number of QNMs
such that $\tau\to 0$  as $a\to M$. For the  extreme black
hole we can use $\hat{\omega} = \omega r_+ = \omega M$, so
the long lived QNMs should be well approximated by
\begin{equation}
\omega_n M \approx {m\over 2} - {1\over 4m}\exp \left[ {\chi -
2n\pi \over 2\delta} + i\zeta \right] \ .
\label{longfreq}
\end{equation} 
This can, of course, be written \cite{sasaki}
\begin{equation}
\omega_n M \approx {m\over 2} - {1\over 4m}e^{(\chi - 2n\pi)/
2\delta} \cos \zeta - {i\over 4m}e^{(\chi - 2n\pi)/ 2\delta}
\sin \zeta \ .
\label{llmodes}
\end{equation}
It can be verified numerically that $\sin\zeta >0$, i.e. these QNMs 
are all damped. 

The approximate result (\ref{llmodes}) provides a useful insight into
the nature of the slowly-damped QNMs, but it is only relevant for very 
nearly extreme black holes. The condition (\ref{modecond}) should, 
however, remain valid for a larger range of spin-rates. 
Thus we can solve this condition numerically to shed light also on the
modes in the near extreme case (that may be astrophysically more relevant).
When we do this we find that the $l=m$ QNMs are qualitatively
similar to (\ref{llmodes}) also for $a \neq M$ in the sense that,
for a give rotation rate, they lie distributed parallel to the 
imaginary $\omega$-axis \cite{modepaper}. As $a \to M$ the separation 
between the mode-frequencies decreases until (\ref{llmodes}) is recovered.


\subsection{The excitation of the slowly damped QNMs}

Having established the existence of the long-lived Kerr QNMs we are ready to turn to our main question: Should we expect these modes to be excited
to a considerable level by a ``realistic'' astrophysical perturbation? 
In order to begin answering this question, we return to equation (\ref{mfield2}). Let us focus on the contribution from the long lived QNMs 
(and the new branch cut that exists in the  $a=M$ case). The remaining contributions to the emerging field should not be much altered from the Schwarzschild case. In particular, all the $l \neq m$ QNMs have frequencies similar to their Schwarzschild counterparts and ought to be excited to a comparable level, leading to a signal that dies out after a few oscillation periods. The contribution from the ``high frequency'' arcs  can be shown, 
just like in the Schwarzschild case \cite{leaver1,na}, to be zero for late-times, and the field that originates from the familiar $\omega=0$ branch cut leads to a late-time power-law tail which has been recently discussed 
by Hod \cite{hod}. For the near extreme case, it follows immediately from 
Cauchy's theorem that the QNM field is 
\begin{equation}
\Phi_{m}(r,\theta,t) \approx \frac{i}{2} \sum_{l=|m|}^{+\infty} \int_{-\infty}^{+\infty} 
dr_\ast^{\prime} \sum_{n} \frac{A^{\rm out}_n}{\alpha_n}e^{-i\omega_n(t-r_\ast -r_\ast^{\prime})}
 S_{l m}(\theta,a\omega_n) {\cal J}_{l m}(r_\ast^{\prime},\omega_n)
\label{qnmfield}
\end{equation}
where we have used the fact that 
$A^{\rm in}(\omega) \approx (\omega -\omega_n) \alpha_n $ near $\omega_n$. 
We have further defined
\begin{equation}
{\cal J}_{l m}(r_\ast^{\prime}, \omega)= \int_{0}^{\pi} d\theta \sin\theta \bar{S}_{l m}(\theta, a\omega)
\Phi_{0}(r_\ast^{\prime},\theta)
\label{initcond}
\end{equation}
Explicit expressions for $A^{\rm in}$ and $A^{\rm out}$ deduced for the 
approximate Teukolsky equation can be found 
in Appendix A. Moreover, the calculation of 
\begin{equation}
\alpha_n= \left. { dA^{\rm in} \over d\omega } \right|_{\omega=\omega_n}
\end{equation}
 is straightforward. 

After some manipulations we obtain for the ``excitation coefficient'' 
\begin{eqnarray}
\frac{A^{\rm out}}{\alpha_n}&=& (-1)^{1/2 +2i\hat{\omega}_n +i\delta_n} (2i\hat{\omega}_n)^{4i\hat{\omega}_n}
\frac{\Gamma(1/2 -2i\hat{\omega}_n -i\delta_n)}{\Gamma(1/2 +2i\hat{\omega}_n -i\delta_n)} \left \{ e^{2\pi \delta_n} - \frac{\Gamma(1/2 -2i\hat{\omega}_n +i\delta_n)\Gamma(1/2 +2i\hat{\omega}_n-i\delta_n)}{\Gamma(1/2 
-2i\hat{\omega}_n -i\delta_n)\Gamma(1/2 +2i\hat{\omega}_n +i\delta_n) } 
\right \} \nonumber \\
&& \times \mbox{LHS}(\omega_n)  
\left [\frac{df}{d\omega} \right]^{-1}_{\omega= \omega_n}
\label{amplit}
\end{eqnarray}
where $\delta_n=\delta(\omega_n)$.
As above $\mbox{LHS}(\omega)$ stands for the 
left-hand side of (\ref{modecond}) 
but without assuming the double limit $(\omega \to m\omega_{+}$, $ a \to M)$. 
The same statement holds for $ f(\omega)= \mbox{RHS}(\omega)-
\mbox{LHS}(\omega)$. 
Its derivative at $\omega= \omega_n$ is
\begin{eqnarray}
\left. \frac{df}{d\omega}\right|_{\omega=\omega_n} &=& i\mbox{RHS}(\omega_n) 
\left \{ \left ( 2r_+ 
+  \delta_n^\prime -\frac{4M}{\sigma_n} \right ) 
\psi(1/2 + 2i\hat{\omega}_n + i\delta_n -4i\tau_n/\sigma_n) \right. 
\nonumber \\  
&-&  \left. \left ( 2r_+  - \delta_n^\prime
-\frac{4M}{\sigma_n} \right ) \psi(1/2 +2i\hat{\omega}_n -i\delta_n -4i\tau_n/\sigma_n) + 2 \left ( \delta_n^\prime \ln(-2i\hat{\omega}_n) + \frac{\delta_n}{\omega_n} \right ) \right \} 
\nonumber \\
\nonumber \\
&-& 2i\mbox{LHS}(\omega_n) \left \{ \delta_n^\prime
\left ( \psi(2i\delta_n) + \psi(-2i\delta_n) + \psi(1+2i\delta_n) + 
\psi(1-2i\delta_n) \right ) \right. 
\nonumber \\
\nonumber \\ 
 &-&  \left. \left ( 2r_+ + \delta_n^\prime \right )
\psi(1/2 -2i\hat{\omega}_n -i\delta_n) + \left ( 2r_+ -
\delta_n^\prime \right ) \psi(1/2 -2i\hat{\omega}_n 
+ i\delta_n) \right \}
\label{fderiv}
\end{eqnarray}
where we have used
\begin{equation}
\delta_n^\prime = \left. { d \delta \over d \omega} \right|_{\omega=\omega_n}
\end{equation}
and
$\psi$ denotes the digamma function \cite{stegun}.
 
Given any reasonable  initial data we can now
 ``reconstruct'' the signal from the long-lived QNMs using 
(\ref{qnmfield}),(\ref{amplit}) and (\ref{fderiv}). 
Results of such a calculation are presented in Figure~\ref{fig5}.  

\begin{figure}[tbh]
\centerline{\epsfysize=6cm \epsfbox{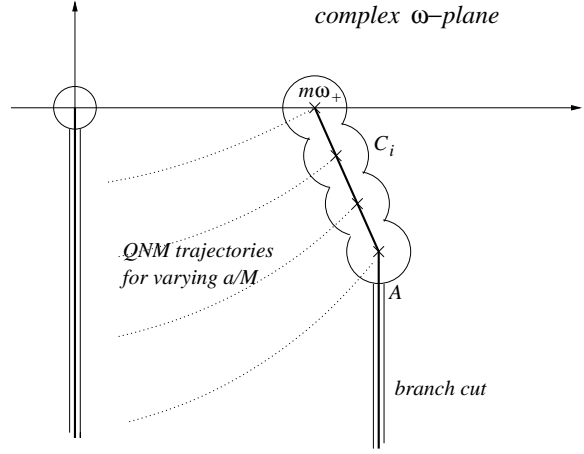}} \caption{Schematic 
description of the QNM location (crosses)
and placement of the required branch cuts (solid lines)
in the case of an extreme Kerr black hole. }
\label{fig3}
\end{figure}

The equations above are, however, somewhat complicated and it is 
useful to consider the $a=M$ case for which they simplify
considerably. Let us first consider the $l=m$ term in (\ref{qnmfield}). 
Bearing in mind that for extreme black holes 
the point $\omega=m\omega_{+}$ is
both a pole and a branch point (see Appendix~B), 
we place the necessary branch cut along the family of Green's function
poles corresponding to (\ref{llmodes}), see Figure~\ref{fig3}. 
After it has  crossed the last mode 
(meaning the mode with the largest imaginary part out of the subset of 
$l=m$ QNMs that approach $\omega=m\omega_+$) 
the cut bends to become parallel to the imaginary 
$\omega$-axis. 
Choosing an integration contour as illustrated in  
Figure~\ref{fig3}, it is straigthforward to show that each of the 
circles $C_i$ will give a contribution equal to the residue of the 
$\omega_i$ pole. That is, we get
\begin{equation}
-\frac{1}{2\pi}\int_{\rm branch-cut}d\omega \frac{e^{-i\omega(t-r_\ast)}}
{2i\omega} S_{mm}(\theta, M\omega) \left ( \frac{A^{out}}{A^{in}} 
{\cal I}_{mm}^{+}(\omega) + {\cal I}_{mm}^{-}(\omega) \right ) = 
S_{\rm M} + {\cal L}_{mm}
\label{bc2}
\end{equation}
where we have defined 
\begin{equation}
S_{\rm M} \equiv \frac{1}{2} \sum_{n=N}^{+\infty} \frac{A^{out}}{\omega_n \alpha_n} 
S_{mm}(\theta,M\omega_n) {\cal I}_{mm}(\omega_n) e^{-i\omega_n(t-r_\ast)}
\label{modesum} 
\end{equation} 
for the mode-sum (with $n$ running from some 
$N>0$ to $+\infty$).
The quantity ${\cal L}_{mm} $ represents the contribution from 
the rest of the branch cut (from point $A$ in Figure~\ref{fig3} downwards).
Explicitly we have
\begin{equation}
{\cal L}_{mm}= \frac{1}{2\pi} e^{-i\omega_{\rm A}(t-r_\ast)}\int_{0}^{-i\infty} d \rho ~ e^{-i\rho(t-r_\ast)} 
\left [ F(\omega_{\rm A} + \rho e^{2\pi i}) - F(\omega_{\rm A} +\rho) \right ] 
\end{equation}
where $\rho= \omega -\omega_{\rm A}$.
The function $F(\omega)$, defined as,
\begin{equation}
F(\omega) \equiv \frac{S_{mm}(\theta, M\omega)}{2i\omega} \frac{A^{\rm out}}{A^{\rm in}}
{\cal I}_{mm}(\omega) \  , 
\end{equation}
can be considered as analytic (after analytic continuation) along 
the integration contour. This will also be true for the function 
$ \tilde{F}(\rho) \equiv F(\omega_{\rm A} +\rho e^{2\pi i}) -
F(\omega_{\rm A} +\rho)$.
Therefore, we can expand this function in a power series around 
$\omega_A$.
\begin{equation}
{\cal L}_{mm}= \frac{1}{2\pi} e^{-i\omega_A(t-r_\ast)}\int_{0}^{-i\infty} 
d \rho ~ e^{-i\rho(t-r_\ast)} 
\left [\tilde{F}(0) + {\cal O}(\rho) \right ] \ . 
\end{equation}    
Integration then yields
\begin{equation}
{\cal L}_{mm} \sim \frac{e^{-\omega_{\rm A}(t-r_\ast)}}{t} \quad \mbox{as} 
\quad  t \to +\infty \ .
\label{bctail}
\end{equation}
As we will soon argue, this contribution can be neglected at late times.

We now compute the mode-sum (\ref{modesum}).
The quantity $\alpha_n$ can be deduced from the derivative of the 
mode-condition (or equivalently, by taking
the $a=M$ limit in (\ref{fderiv}) ). We  get
\begin{equation}
\left. {df \over d\omega} \right|_{\omega=\omega_n}
\approx -8imM\delta_n \exp\left[ \left( i - {1\over 2\delta_n} \right)
( \chi-2n\pi) + i(2i\delta_n-1)\zeta \right] \ .
\end{equation}
Given this, we find that
\begin{equation}
 { A^{\rm out} \over  \alpha_n } \approx (-1)^{1/2 +im +i\delta_n} (im)^{2im} 
\frac{ \Gamma(1/2 -im -i\delta_n)}{\Gamma(1/2 +im -i\delta_n)}qe^{i\chi}
\left [ e^{2\pi\delta_n} - \frac{\Gamma(1/2 -im +i\delta_n)
\Gamma(1/2 +im -i\delta_n)}
{\Gamma(1/2 -im -i\delta_n)\Gamma(1/2 +im +i\delta_n)} \right ] 
\left[ {df \over d\omega} \right]_{\omega=\omega_n}^{-1} \ .
\label{amplit2}
\end{equation}
Hence, we have shown that
\begin{equation}
 { A^{\rm out} \over  \alpha_n } \sim e^{-n\pi/\delta_n}
 \label{longcoeff}
\end{equation}
i.e. each long-lived QNM is excited to an exponentially small
amplitude.
The total QNM response,  $S_{\rm M}$, can now be approximated by
\begin{equation}
S_{\rm M} \sim \sum_{n=N}^{+\infty} \frac{A^{\rm out}}{\alpha_n} 
e^{-i\omega_n(t-r_\ast-r_{\ast}^{\prime})} \ .
\end{equation}
Using (\ref{amplit2}) and (\ref{llmodes}), we see that,
\begin{equation}
S_{\rm M} \sim \sum_{n=N}^{+\infty} \exp\left[- (1-2i\delta){n\pi\over \delta}
-\beta e^{-n\pi/\delta} \right] \ ,
\end{equation}
where $\beta = iu\exp(\theta/2\delta-i\varphi)/4mM$, and we have
 introduced the retarded time $u= t- r_\ast -r_\ast^\prime$.
This can be  written as
\begin{equation}
S_{\rm M}  \sim \sum_{n=0}^{+\infty} \exp\left[- (1-2i\delta){n\pi\over \delta}-\tilde{\beta} e^{-n\pi/\delta} \right] \ ,
\end{equation}
with  $\tilde{\beta}= \beta e^{N\pi/\delta}$. 
Since the main contribution to the sum is coming from the large values of $n$, 
we can approximate the sum by an integral,
\begin{equation}
S_{\rm M} \sim {\delta \over \pi} \int_0^\infty \exp \left[ -\alpha x -
\tilde{\beta} e^{-x} \right] dx
\end{equation}
where $\alpha = 1-2i\delta$. Expanding the expression in the bracket
as a Taylor series and integrating we get
\begin{equation}
S_{\rm M} \sim {\delta \over \pi} \sum_{k=0}^{\infty} 
{ (-1)^k \tilde{\beta}^k \over k! (k+\alpha) } = {\delta \over \pi} \tilde{\beta}^{-\alpha}
\gamma(\alpha, \tilde{\beta}) \ ,
\end{equation}
after identifying the series representation of the incomplete
gamma function \cite{stegun}. 
Since $\gamma(\alpha, \tilde{\beta}) \to \Gamma(\alpha)$ as $\tilde{\beta} 
\to \infty$ we can deduce that
\begin{equation}
S_{\rm M} \sim {e^{-im\omega_{+}t} \over t} \quad \mbox{ as } t \to \infty \ .
\label{qtail}
\end{equation}
Since this decays slower than (\ref{bctail}) we can neglect the contribution
from the extreme Kerr branch cut.

Next, we turn to the $l>m$ terms in (\ref{mfield2}).
In view of the absence of any poles in the vicinity of
 $\omega = m\omega_{+}$, we simply
place the branch cut parallel to the imaginary axis. 
The only multivalued quantity at $\omega=m\omega_{+}$ in the 
integrand of (\ref{mfield2}) is the ratio $A^{\rm out}/A^{\rm in}$ 
(see Appendix A). A standard branch cut calculation then yields
\begin{eqnarray}
\Phi_{m}(r,\theta,t)&=& \frac{1}{4\pi M}\int_{-\infty}^{+\infty} dr_\ast^\prime e^{-imu/2M} 
\int_{0}^{-i\infty} d\tau e^{-i\tau/M} S_{l m}(\theta, a\tau/M + am/2M) 
{\cal J}_{l m}(r_\ast^\prime, \tau/M +m/2M)  
\nonumber \\
&& \times \left [ \frac{A^{\rm out}}{A^{\rm in}}(\tau e^{2\pi i}/M + m/2M) -\frac{A^{\rm out}}{A^{\rm in}}(\tau/M +m/2M) \right ]
\label{bclneqm}
\end{eqnarray}
We can split the frequency integral into two parts (for brevity we omit
the integrands):
\begin{equation}
\int_{0}^{-i\infty} d\tau \left \{ ... \right \}=
 \int_{0}^{-i\epsilon} d\tau \left \{ ... \right \}  + \int_{-i\epsilon}^{-i\infty} d\tau \left \{ ... \right \}  
\label{bc2lm} 
\end{equation}
The constant $\epsilon$ is chosen in order to
allow us  to use the $a=M$, $\omega \approx m\omega_{+}$ approximation 
\begin{eqnarray}
\frac{A^{\rm out}}{A^{\rm in}}(\omega,a=M) &=& (-1)^{1/2 +2i\omega M +i\delta} (2i\omega M)^{4i\omega M} \frac{ \Gamma(1/2 -2i\omega M -i\delta)}{\Gamma(1/2 +2i\omega M -i\delta)}
\nonumber \\
&\times&  \left [ \frac{\Gamma(2i\delta) \Gamma(1+2i\delta) \Gamma(1/2 +2i\omega M -i\delta)
\Gamma(1/2 -2i\omega M -i\delta)}{ \Gamma(-2i\delta) \Gamma(1-2i\delta) \Gamma(1/2 -2i\omega M
+i\delta) \Gamma(1/2 +2i\omega M +i\delta) } + (8\omega M\tau)^{2i\delta}
 \right ]
 \nonumber \\ 
 &\times &\left [ { \Gamma(2i\delta) \over \Gamma(-2i\delta)} {\Gamma(1+2i\delta) \over 
\Gamma(1-2i\delta) } \left ( {\Gamma(1/2-2i\omega M-i\delta) \over
\Gamma(1/2-2i\omega M + i\delta) } \right)^2 + (-8\omega M \tau)^{2i\delta} \right ]^{-1}
\label{exratio} 
\end{eqnarray} 
The key point here  is that for $l \neq m$ the quantity $\delta$ is 
almost purely imaginary near  $m\omega_{+}$, with positive imaginary part. 
We then have for $\tau \to 0$;
\begin{equation}
\tau^{2i\delta} \to +\infty
\end{equation}
Hence, the expression inside the bracket in the first integral 
of the right-hand side of (\ref{bc2lm}), will be vanishingly 
small and therefore 
the contribution from the part of the branch 
cut close to $m\omega_{+}$ is negligible for $l \neq m$. 
The second integral in (\ref{bc2lm}) can be 
evaluated using arguments similar to the ones involved 
in the derivation of (\ref{bctail}). We end up with 
an exponentially decaying field that can be neglected compared
to (\ref{qtail}).

Finally, we have to consider the contribution from the 
remaining QNMs. But because all these modes are ``short-lived''
their contribution is important only during the early phase of
QNM ringing, cf. the Schwarzschild results \cite{na}.

The main conclusion of this Section is that
even though the contribution of each long lived QNM 
to the emerging field is exponentially small, the total mode 
contribution cannot be neglected. When summed, 
the slowly damped QNMs of an extreme Kerr black hole 
give rise to  an oscillating 
signal whose magnitude falls off with time as a power-law. 
This result is particularly interesting since 
the decay of 
this signal is considerably slower than the standard power-law tail
in the Kerr case 
\cite{hod,barack1,barack2}.
It is  worth  mentioning that oscillating power-laws are 
known to arise in standard scattering theory
whenever the Green's function has multiple poles \cite{newton}. In our case, 
one could argue that it is the exponentially small ``spacing'' between 
neighbouring poles (see expression (\ref{llmodes})) that causes
the $1/t$ power-law.

It is also worth pointing out that
the prediction (\ref{longcoeff}) for the excitation of each 
individual long-lived QNM  agrees with the intuitive arguments we made in the 
Introduction. For example, we see that the leading QNM 
is {\em not} excited for $a=M$ (as it corresponds to the limit 
$n \to \infty$ in (\ref{longcoeff})). 
This is in agreement with, and generalises, the 
$ l=m \gg 1 $ result (which is valid only for the fundamental mode) 
 obtained in \cite{ferrari} using the eikonal approximation.    

\section{Numerical results}

\subsection{Time evolutions}

In the previous Section we used analytic approximations
to  analyze the long-lived QNMs of a
near extreme Kerr black hole. We obtained results that agree 
with our anticipations
that these modes would not individually be excited to a significant level. 
But our calculation also provided surprises, the most important being that
the large number of slowly damped QNMs combine to give a significant
signal the amplitude of which falls of as $1/t$ at late times. 
 However, in view of the many approximations involved in the 
derivation of (\ref{qtail})  considerable 
caution is warranted, and a confirmation of the analytic prediction is 
desirable. One way to obtain support for the results would be to 
perform time-evolutions of the Teukolsky equation from given initial data.
In other words, the recent effort to develop a framework for doing
perturbative time-evolutions for Kerr black holes 
\cite{tcode1,tcode2,tcode3,tcode4} provides the means for testing our 
analytical predictions. 

\begin{figure}[tbh]
 \centerline{\epsfysize=5cm \epsfbox{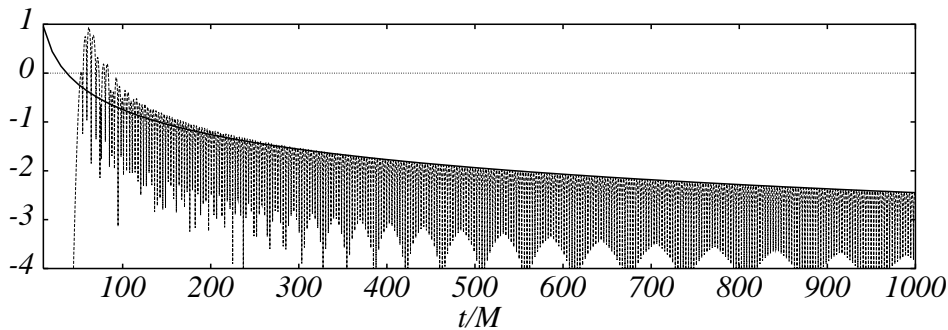}}
\caption{A numerical evolution showing the late-time behaviour 
of a scalar field in the  geometry of a rapidly rotating Kerr
black hole. We show (on a logarithmic scale) the field as viewed by an observer situated well away from the black hole for $a=M$. At late times the 
field falls off according to an oscillating power-law with the amplitude decaying as $1/t$. The data 
corresponds to a narrow Gaussian (initially centered at $r_\ast=50M$) that hits the black hole, and is observed at $r_\ast=10M$ }
\label{fig4}
\end{figure}

We have performed a set of evolutions 
(for various values of $m$) using the same scalar field code that 
was used to study superradiance  in a dynamical 
context \cite{tcode4}.
As initial data we have chosen a generic Gaussian pulse originally 
located far away from the black hole. The evolution of this pulse, as it 
travels towards the black hole and excites the QNM ringing is then studied.
The numerical results we obtain support the following conclusions:
i) In the case of extreme Kerr black holes ($a=M$)
we verify the predicted oscillating $1/t$ behaviour for all $m\neq 0$,
cf. Figure~\ref{fig4}.
ii) For $a<M$ we recover the anticipated exponential fall-off at late times, 
cf. \cite{tcode2}. Still, the emerging signal differs considerably from a
single QNM oscillation at intermediate times for near extreme black holes. 
This agrees well with the analytical results from the
previous Section: For near extreme black holes a large number of 
long-lived QNMs are excited to roughly the same level,  
and for a considerable time window the resulting signal is a 
superposition of many slowly decaying exponentials (each with a very small
amplitude). 
As we depart further from $a=M$ we retain the standard result: The signal 
is completely dominated by the slowest damped QNM. 
iii) For axisymmetric perturbations ($m=0$) the numerical evolution
recovers the standard power-law tail. For our particular choice of
initial data (that contains the $l=0$ multipole) the tail
falls off as $t^{-3}$. This agrees with the predictions of, for example, 
Ori and Barack \cite{barack1,barack2}.

Although our numerical simulations generally support the analytic results, 
there is still room for some caution. It is very difficult to investigate the 
late-time behaviour of a perturbed Kerr black hole using the Teukolsky code. 
After all, we would like to be able to distinguish a slowly damped
exponential from an oscillating power-law at very late times. 
Given our current
numerical code we cannot obtained absolute proof of our analytic results. 
Detailed convergence tests show that the code has the 
expected properties, but also unveil that it is difficult to reliably 
determine the late time oscillating tail predicted by our analytical 
work. While the overall amplitude envelope can be determined for 
many hundred dynamical timescales as the resolution is increased, the signal
typically goes out of phase on a shorter timescale. This means that, even 
though the numerical results provide support for our theoretical 
predictions, one would have to build an evolution 
code that could be trusted at very late times, eg. based on  a double-null
evolution, in order to obtain definite results. 

\begin{figure}[tbh]
\centerline{\epsfxsize=7cm \epsfbox{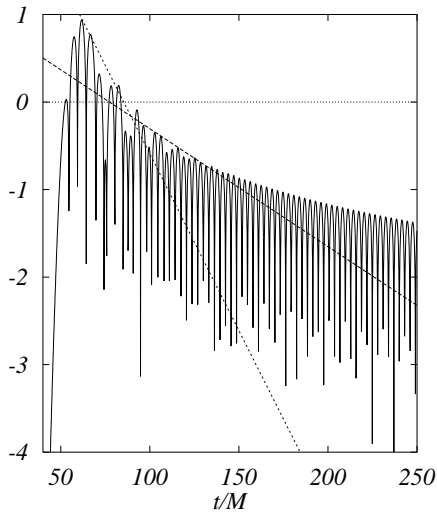}} 
\caption{We compare the late-time QNM oscillations in a typical scalar
field evolution (initial data and observer location are the same as in 
Figure~\ref{fig4}) for $m=2$ and
$a=0.99M$ to the damping rates of the two slowest damped
QNMs (indicated by dashed lines). From this figure it is clear that the late-time behaviour 
is not well represented by either of these modes individually.}
\label{twomode}
\end{figure}


\subsection{Reconstructing the long-lived QNM signal}

In addition to investigating the late-time field through direct 
numerical evolutions, we can reconstruct it from our analytical
results.  Using expressions (\ref{qnmfield}) - (\ref{fderiv}) 
it is a straightforward task to calculate the long-lived QNM
signal provided that we have already acquired the corresponding 
mode-frequencies. 
In this way one can expect to obtain a fairly accurate description 
of the full signal for sufficiently late times (i.e. after 
some e-folding times of the dominant short-lived QNM: typically 
the slowest damped $m=-l$ mode), spanning a time window 
of several hundred dynamical timescales. Eventually one would expect this 
signal (after all, it decays exponentially) to give way to the 
standard  power-law tail. However, as is clear from all our illustrations
this will not happen until at very very late times.

\begin{figure}[tbh]
\centerline{\epsfysize=6cm   \epsfbox{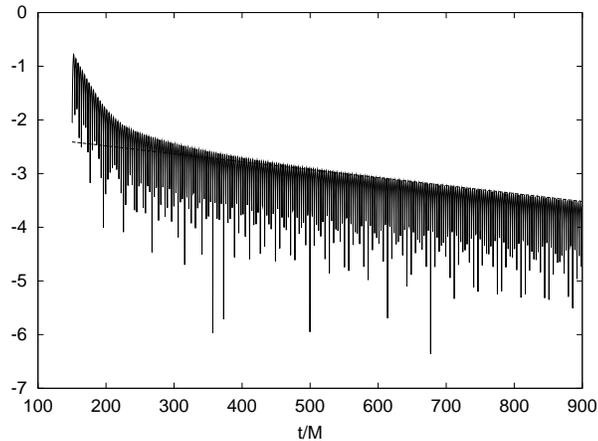}}
\caption{The signal (shown on a logarithmic scale)  from the long-lived QNMs 
(the first ten modes are included here), for an $a=0.9999M$ Kerr black hole. 
The required mode-data is given in Table~\ref{tab1}. 
The observer is located at $r_\ast= 100M, \theta= \pi/2 $ and a narrow
Gaussian pulse (centered at $r_{\ast}=50M$) was used as initial data.
The dashed line represents the damping rate of the slowest decaying QNM. 
It is clear that for several hundreds of black hole's dynamical times, the 
slowest damped mode is not dominant. 
This is in contrast to the results for slowly
rotating black holes.}
\label{fig5}
\end{figure}

In Table~\ref{tab1} we present a small subset of our numerical data for
long-lived QNMs frequencies and the corresponding 
excitation coefficents for near extreme black holes. 
As anticipated, these amplitudes vanish 
as $a \to M$. Moreover, as already pointed out, the higher overtones are 
excited to a level comparable to  the leading mode. 
In  Figure~\ref{fig5} we illustrate the typical 
field as constructed from the long-lived QNMs. 
For this particular calculation, we have
chosen, as initial data, a Gaussian pulse with angular dependence 
$\sim Y_{22}(\theta,\phi)$. It follows then from (\ref{qnmfield}), that
only the $l=m=2$ field component will be important at late times.
The result in Figure~\ref{fig5} should be compared to the fully
numerical evolution data in Figures~\ref{fig4}-\ref{twomode}. 

\section{A physical interpretation: The superradiance resonance cavity}

\begin{figure}[tbh]
\centerline{\epsfxsize=8cm \epsfbox{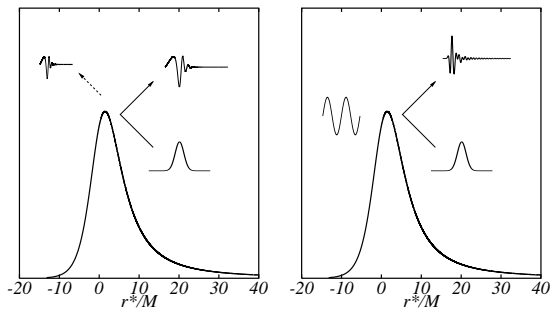}}
 \caption{Schematic explanation of the new phenomenon seen in the numerical evolutions of Kerr perturbations. The left panel illustrates the standard scenario: An infalling pulse excites the QNMs that then propagate to infinity 
and the horizon. At late times, backscattering due to the curvature in the far-zone dominates and leads to the familiar power-law tail behaviour. 
Right panel: For frequencies that i) lie in the superradiant regime, and ii) experience a ``potential peak'' in the region $[r_+,\infty]$ there will be a superradiance resonance cavity outside the black hole. At late times, the 
waves leaking out of this cavity to infinity dominate the signal.}
\label{fig6}
\end{figure}

\begin{figure}[tbh]
\centerline{\epsfxsize=10cm \epsfbox{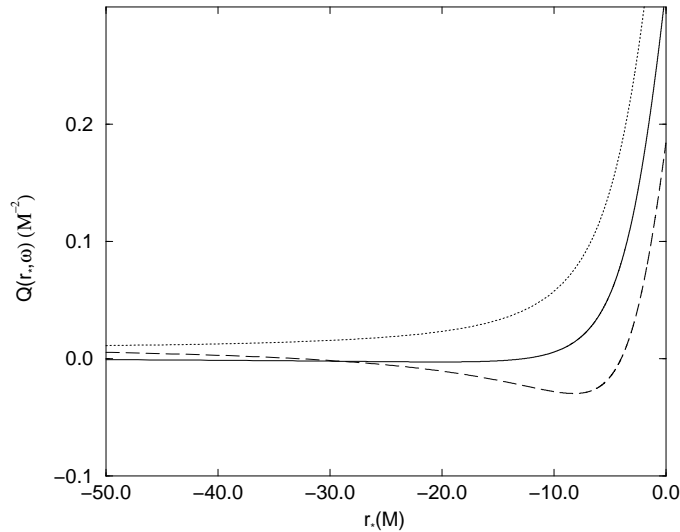}}
\caption{Graph of the $l=m=2$ effective radial potential
$Q(r_{\ast},\omega)$ as a function of $r_{\ast}$, for $\omega= 0.9m\omega_{+}$ (dashed curve), $m\omega_{+}$ (solid curve) , $1.1 m\omega_{+}$ (dotted curve). The black hole spin is $a=0.999M$. Note the appearance of a potential barrier
($Q < 0$) as soon as we enter the superradiant frequency regime 
$\omega<m\omega_+$.}
\label{fig7}
\end{figure}

\begin{figure}[tbh]
\centerline{\epsfysize=8cm \epsfbox{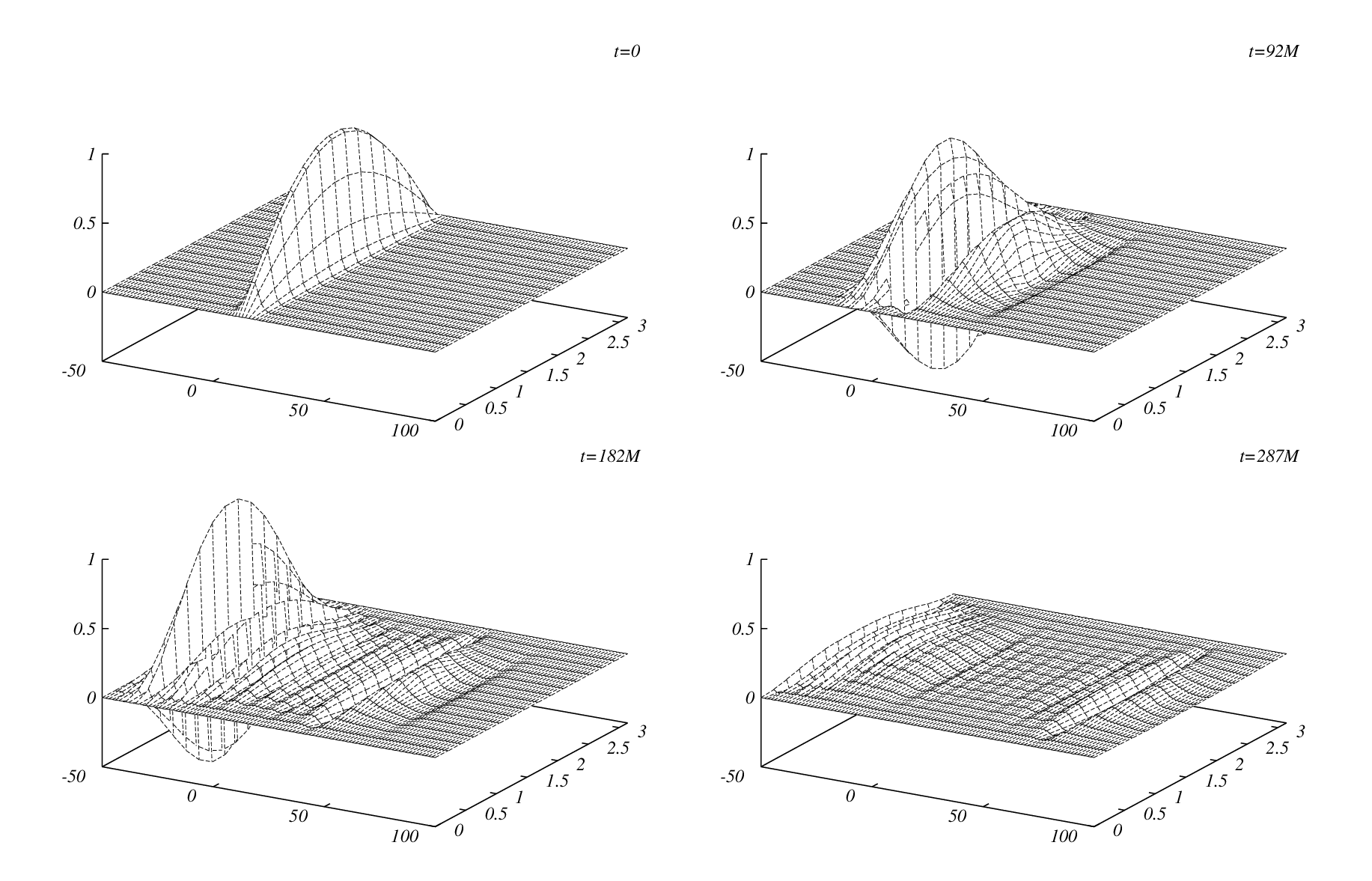}}
\centerline{\epsfysize=8cm \epsfbox{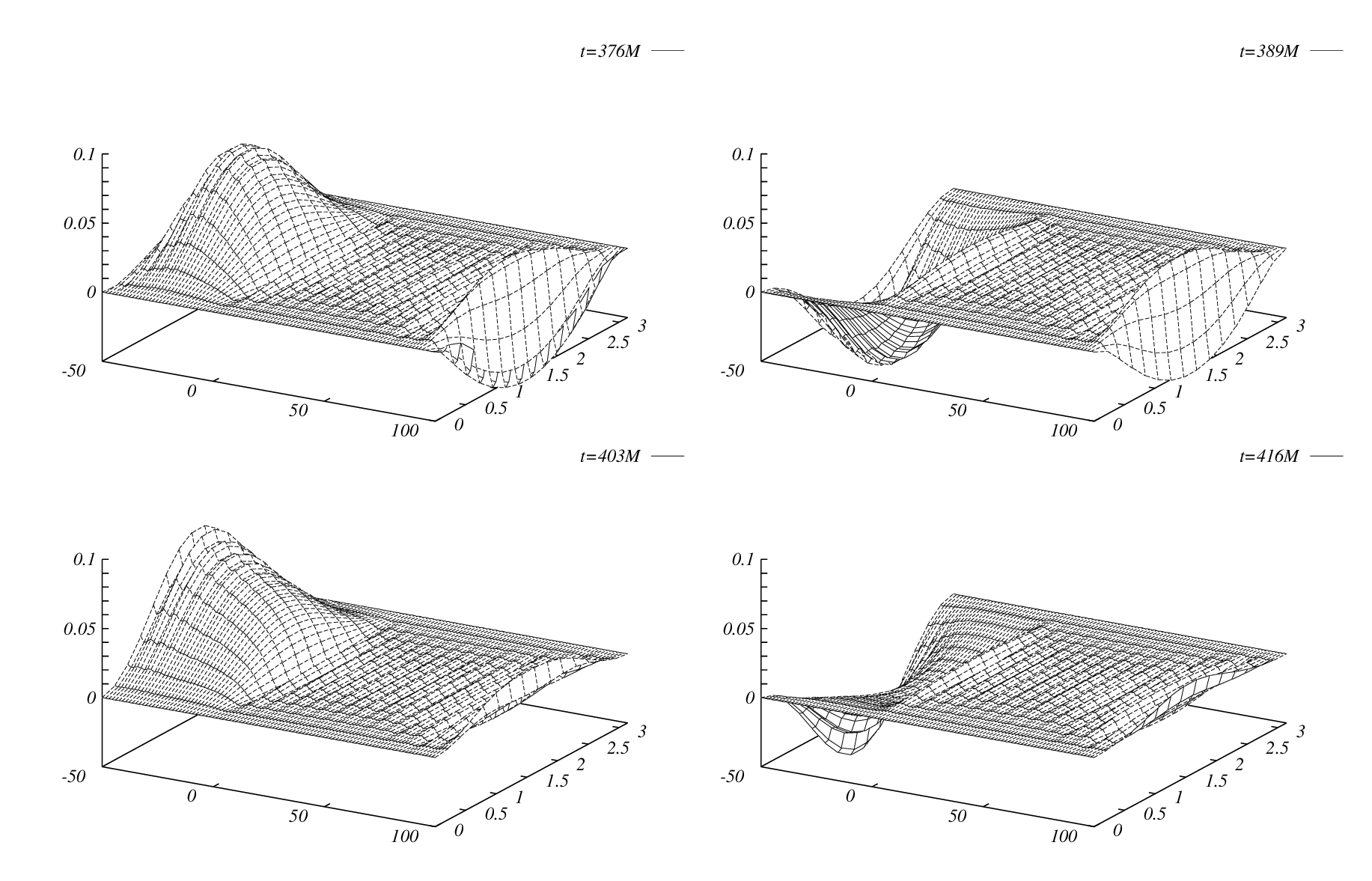}}
\caption{ A series of snapshots of a typical numerical evolution of a
rapidly spinning black hole. We show the scalar field a function of both $r_\ast$ and $\theta$ at various times (as indicated in the upper right hand corner
of each image). The images illustrate how ``trapped'' oscillations
develop in the  vicinity of the black hole's horizon. After an 
initial drop in amplitude, cf. the first four images, 
(roughly at the timescale of the e-folding time of 
the rapidly damped ``Schwarzschild-like'' QNMs) these oscillations
decay very slowly. These last four snapshots indicate the presence of a 
very slowly damped standing wave
in the region just outside the horizon.}
\label{fig8}
\end{figure}

Our analytical results, supported by the outcome of numerical evolutions, 
give rise to an intriguing picture. It would seem as if the
extremely long-lived QNMs will, even though  they are 
hardly excited at all individually,  dominate the signal from a very rapidly
spinning black hole. The emergence of this late-time signal, to which
many QNMs contribute would be a new phenomenon in black-hole physics.
Given that both the standard QNMs and the power-law tail have simple 
intuitive explanations it may be worthwhile trying to understand 
the extreme Kerr case in a similar way. 
From the evidence provided by our numerical evolutions, see Figure~\ref{fig8}, we propose the following ``explanation''.
Consider the fate of an essentially monochromatic wave that falls 
onto the black hole, cf. Figure~\ref{fig6} for a schematic description. 
Provided that the 
frequency is in the interval $0< \omega < m\omega_+$  
the wave will be superradiant: scattering of these waves in the black hole's
ergosphere results in their amplification by extraction of the black hole's 
rotational energy. In effect, this means that a distant  observer will 
see waves ``emerging from the horizon'', cf. (\ref{in}), even though 
a local  observer sees the waves crossing the event 
horizon (at $r_+$) \cite{teuk2}. In addition 
to this, one can establish that the effective potential has a 
peak outside the black hole (which is not immediately obvious since 
the  ``potential'' $Q$ is frequency dependent in the Kerr case) for a 
range of frequencies including the superradiant interval. Indeed, 
we can find  the following 
approximate expression
for the potential near the horizon (valid for $a=M$ only)
\begin{equation}
Q(r,\omega)= (\omega -m\omega_{+})^2 -2(\omega-m\omega_{+})\frac{m}
{r_\ast} + {\cal O} \left ( \frac{\ln(r_{\ast}/M)}{r_\ast^2} \right )
\end{equation}
This shows that, for $\omega \approx m\omega_{+}$, there will be a peak 
(corresponding to a minimum of $Q$) just 
outside the horizon. This can be verified  by graphing the exact potential (\ref{potent}), cf. Figure~\ref{fig7}.     

The combination of the causal boundary condition at the horizon 
effectively corresponding to waves  ``coming out of the black hole'' 
(according to a 
distant observer) and the presence of a potential peak  leads to 
waves potentially being trapped in the region close to the horizon. 
In effect, 
there is a ``superradiance resonance cavity'' outside the black hole.  
Again according to a distant observer, waves can only escape from 
this cavity by leakage through the potential barrier to infinity. 
Since the superradiant amplification is strongest for frequencies 
close to $m\omega_+$, waves in the cavity experience a kind of 
parametric amplification and at very late times the 
dominant oscillation frequency ought to be $m\omega_+$. This is
exactly what we have deduced from  our analytic and numerical calculations.
In Figure~\ref{fig8} we present a series of snapshots of the 
evolution of a scalar field around an extreme Kerr black hole. In this series
of pictures one can see how ``trapped'' oscillations
develop in the  vicinity of the black hole's horizon. After an 
initial drop in amplitude (roughly at the timescale of the e-folding time of 
the rapidly damped ``Schwarzschild-like'' QNMs) these oscillations
decay very slowly. These snapshots indicate the presence of a standing wave
in the region just outside the horizon. In the extreme black hole case leakage from the superradiance cavity leads to the observed $1/t$ decay. In the near extreme case, the existence of the cavity provides an intuitive 
explanation for the extremely slow damping of corotating QNMs 
with frequencies close to $m\omega_+$. In a way, the superradiance cavity
we have described here can be viewed as the black hole analogue of the 
potential ``well''  present in the gravitational field of ultracompact stars \cite{chandra2} which also leads to the existence of a long-lived family
of $w$-modes \cite{ultra}.  

\section{Concluding Discussion}

We have presented the results of an investigation into the 
late-time behaviour of a perturbed Kerr black hole. An analytic
calculation for scalar fields in the geometry of an extreme Kerr 
black hole provided two important results. The first concerns the 
level of excitation of the QNMs that
become very slowly damped as $a\to M$. We find that these
modes are much more difficult
 to excite than their rapidly damped 
Schwarzschild counterparts. This means that (individually)
 these modes may not be easy to detect with the new generation of  
gravitational-wave detectors. 
However, there may still be a detectable, slowly damped, QNM signal
owing to the fact that a large number of virtually undamped QNMs 
exist for each value of $m\neq 0$. We find that these modes combine 
in such a way that the field oscillates with an amplitude that 
decays as $1/t$ at late times. 
This decay is considerably slower than the standard power-law tail. 
We have used numerical time-evolutions of the Teukolsky  equation to
verify this analytic prediction for extreme 
black holes. 

After extending our study to near extreme black holes we 
find that, even though the QNM signal decays exponentially in the 
familiar way, there is still a considerable time interval where a large 
number of small-amplitude slowly-damped QNMs are present in the signal.
This means that the signal from a rapidly spinning black hole would not be 
well represented by a single QNM approximation.  
Only at very late times do the leading mode become dominant and eventually
it should give way to the power-law tail  predicted in previous work
\cite{barack1,barack2,hod}. 

Whether our results are astrophysically relevant or not is in many ways still 
an open question. In particular, the excitation of the long-lived QNMs 
by more realistic initial data (representing, for example, the close-limit
approximation of merging Kerr black holes) must be studied. 
One should obviously also try to understand whether one would expect
to find almost extreme black holes in Nature (recall that Thorne \cite{kip}
has shown that
accretion cannot spin a black hole up beyond $a=0.998M$).
It seems quite reasonable, however, to expect that the effects we have 
observed will play a role at intermediate times for 
rapidly rotating
non-extreme black holes.
If our predictions are correct one must investigate in detail to what 
extent the late-time signals from a rapidly rotating black-hole are 
detectable even though they contain a large number of QNMs (each with a 
small amplitude). In particular, it is crucial to determine whether 
one can hope to infer the black-hole parameters from a signal that 
contains such  a superposition of QNMs (perhaps using techniques similar 
to those introduced in \cite{nollert}). 
It is also interesting to ask whether there exists a critical value of 
the rotation parameter $a$ above which the new effect we have observed 
becomes relevant (recall that our approximate modes are only relevant 
for $a\approx M$). More detailed numerical work is needed to
answer this question.
Finally, we note that the observed phenomenon can be intuitively explained
in terms of a ``superradiance resonance cavity'' outside the black hole, 
in which waves of certain frequencies are effectively trapped.
 As the waves slowly leak out from this cavity, they give rise the 
``long-lived'' QNM signal observed at infinity. In conclusion, it is 
interesting to note that, even though black-hole perturbations is a
very well researched field, it can still provide interesting and
surprising results. 


\acknowledgments

The authors would like to thank Misao Sasaki for helpful discussions. 
K.G. thanks the members of the Cardiff Relativity Group for useful discussions,  and the State Scholarships Foundation of Greece for financial support. 
N.A. is a Philip Leverhulme Prize Fellow, and also acknowledges support from PPARC via grant number PPA/G/1998/00606 and the European Union via the network ``Sources for Gravitational Waves''.


\appendix

\section{Approximate solutions of the Teukolsky equation for near
extreme rotation}

In this Appendix we reproduce the solution of the Teukolsky equation for $a\approx M$, 
$\omega \approx m\omega_{+}$, as given by Teukolsky and Press \cite{teuk2}. 
Throughout this paper we have (as far as possible) used the notation of 
{\em their} Appendix A1. The equation to be solved is not the ``usual'' Teukolsky equation, but the one that follows if we consider the problem in ``ingoing Kerr'' coordinates. These are related to the Boyer-Lindquist coordinates by
\begin{eqnarray}
v &=& t + \int { r^2 + a^2 \over \Delta} dr \ , \\ \tilde{\varphi}
&=& \varphi + \int { a\over \Delta} dr \ . \\
\end{eqnarray}
In these coordinates the separation of variables follows from (note that the radial functions here differ from the one used in Section IIA by the factor $(r^2+a^2)^{-1/2}$)
\begin{equation}
\Phi_{l m} = \int \tilde{R}_{l m}(\omega,r)S_{l m}(\theta,a\omega)
e^{im\tilde{\varphi} - i\omega v} d\omega \ .
\end{equation}
Comparing this expression to the standard Boyer-Lindquist decomposition
\begin{equation}
\Phi_{l m} = \int R_{l m}(\omega,r)S_{l m}(\theta,a\omega) 
e^{im\varphi- i\omega t}d\omega \ ,
\end{equation}
we can immediately relate the radial wavefunctions via
\begin{equation}
R_{l m}(\omega,r) = \tilde{R}_{l m}(\omega,r) \exp \left[ -i \int {K \over
\Delta} dr \right] \ .
\label{connect}
\end{equation}
Since
\begin{equation}
\int {K \over \Delta} dr \sim  \left\{ \begin{array} {ll} \omega
r_\ast \ , \quad \mbox{ as } r_\ast \to +\infty \ , \\
kr_\ast \ , \quad \mbox{ as } r_\ast \to -\infty \ .
\end{array} \right.\\
\end{equation}
where we recall that $k=\omega -m \omega_+$. Hence, we see that the 
``physically acceptable'' solution behaves as
\begin{equation}
\tilde{R}^{\rm in} \sim \left \{ \begin{array}{ll} 1 \quad \quad \quad 
\mbox{as} \quad r\to r_+ \ , \\   
Z^{\rm out} r^{-1} e^{2i\omega r_\ast} + Z^{\rm in} r^{-1}  
\quad \mbox{as } r\to +\infty \ .
\end{array} \right.
\label{apin}
\end{equation}
In ingoing Kerr coordinates, the radial Teukolsky equation takes the form,
\begin{equation}
x(x+\sigma)\frac{d^2\tilde{R}_{l m}}{dx^2} -\{2i\hat{\omega}x^2 + 2x(2i\hat{\omega} -1) -\sigma +4i\tau\}\frac{d\tilde{R}_{l m}}{dx} -\{2i\hat{\omega}(x+1) +\lambda\}\tilde{R}_{l m}=0
\label{ingteuk}
\end{equation}
The variables $x$ etcetera were defined by (\ref{ptvar1})-(\ref{PTvars}).
The double limit $a\to M$, $\omega \to m\omega_{+}$ 
corresponds to $\sigma \to 0$,
$~\tau \to 0$. Let us first consider this equation in the limit when $x>>\mbox{max }(\sigma,\tau)$, i.e., for large radii. Then (\ref{ingteuk}) is 
well approximated by
\begin{equation}
x^2 \frac{d^2\tilde{R}_{l m}}{dx^2} - \{ 2i\hat{\omega}x^2 +
2x(2i\hat{\omega} -1 ) \} \frac{d\tilde{R}_{l m}}{dx} - \{2i\hat{\omega}
(x+1) +\lambda \} \tilde{R}_{l m} = 0 \ .
\label{ingteuk1}
\end{equation}
A solution to (\ref{ingteuk1}) that satisfies (\ref{apin}) can be written in terms of confluent hypergeometric functions \cite{stegun}
\begin{equation}
\tilde{R}_{l m}= Ax^{ -1/2 +2i\hat{\omega} +i\delta} M(1/2  +2i\hat{\omega} +i\delta, 1 +2i\delta,2i\hat{\omega} x) + B(\delta \rightarrow -\delta)
\label{largex}
\end{equation}
where $A$,$B$ are constants and the notation $ (\delta \rightarrow -\delta)$
means ``replace $\delta$ by $-\delta$ in the preceding term''. 
We next turn to the case when $x<<1$, i.e., try to find a solution that is 
valid close to the black hole's horizon. The original equation simplifies (slightly) to
\begin{equation}
x(x+\sigma)\frac{d^2\tilde{R}_{l m}}{dx^2} - \{ 2x(2i\hat{\omega} -1) 
-\sigma + 4i\tau)\} \frac{\tilde{R}_{l m}}{dx}
 -\{ 2i\hat{\omega}+\lambda \} \tilde{R}_{l m} = 0 \ .
\end{equation}
This is the hypergeometric equation, and one solution can be written as
\begin{equation}
\tilde{R}_{l m} = {}_{2}F_{1}(1/2 -2i\hat{\omega}+i\delta, 1/2-2i\hat{\omega}-i\delta, 1- 4i\tau/\sigma,
 -x/\sigma) \ .
\label{smallx}
\end{equation} 
It is straightforward to verify that ${}_{2}F_{1}\to 1$ as $x \to 0$, 
which means that this solution has the desired ``purely ingoing wave'' 
behaviour close to the event horizon. The solutions (\ref{largex})
and (\ref{smallx})  can be matched in the overlap region, 
max($\sigma$,$\tau$) $\ll x \ll 1$. 
The $x \to 0$ limit of (\ref{largex}) yields,
\begin{equation}
\tilde{R}_{l m} \to A x^{-1/2 +2i\hat{\omega} +i\delta} + B(\delta \to -\delta)
\label{largex2} 
\end{equation}
Similarly, for $x \to \infty$, (\ref{smallx}) becomes,
\begin{equation}
\tilde{R}_{l m} \to \frac{ \Gamma(1 -4i\tau/\sigma) \Gamma(2i\delta) \sigma^{1/2 -2i\hat{\omega} 
-i\delta} }{ \Gamma(1/2 -2i\hat{\omega} +i\delta) \Gamma( 1/2 +2i\hat{\omega} +i\delta +4i\tau/\sigma) } x^{-1/2 +2i\hat{\omega} +i\delta} + (\delta \to -\delta)
\label{smallx2}
\end{equation} 
We can extract $A$ and $B$ by matching the solutions (\ref{largex2}) and
(\ref{smallx2}), 
\begin{eqnarray}
A &=& \frac{\Gamma(1 -4i\tau/\sigma) \Gamma(2i\delta) \sigma^{1/2 -2i\hat{\omega} -i\delta}}
{\Gamma(1/2 -2i\hat{\omega} + i\delta) \Gamma(1/2 +2i\hat{\omega} + i\delta -4i\tau/\sigma) }
\label{A}
\\
B &=& \frac{\Gamma(1 -4i\tau/\sigma) \Gamma(-2i\delta) \sigma^{1/2 -2i\hat{\omega} +i\delta}}
{\Gamma(1/2 -2i\hat{\omega} - i\delta) \Gamma(1/2 +2i\hat{\omega} - i\delta -4i\tau/\sigma) }
\label{B}
\end{eqnarray}
On the other hand, approximating (\ref{largex}) for $x \to \infty$ we get for the amplitudes $ Z^{\rm in}$,$Z^{\rm out}$,
\begin{eqnarray}
Z^{\rm in}&=& r_{+} A \frac{\Gamma(1+2i\delta)}{\Gamma(1/2 -2i\hat{\omega} +i\delta)}
(-2i\hat{\omega})^{-1/2 -2i\hat{\omega} -i\delta} + B(\delta \to -\delta) \\
Z^{\rm out} &=& r_{+} A \frac{\Gamma(1+2i\delta)}{\Gamma(1/2 +2i\hat{\omega} + i\delta)}
(2i\hat{\omega})^{-1/2 +2i\hat{\omega} -i\delta} + B(\delta \to -\delta)
\end{eqnarray}
Using (\ref{connect}), it is easy to see that $A^{\rm in}= Z^{\rm in}$ and $A^{\rm out} =Z^{\rm out} $ where $A^{\rm out}$ and $A^{\rm in}$ are the 
asymptotic amplitudes in the standard Boyer-Lindquist decomposition.
Explicitly we get
\begin{eqnarray}
A^{\rm in}(\omega)&=& r_{+}(-2i\hat{\omega})^{-1/2 -2i\hat{\omega} -i\delta} \sigma^{1/2 -2i\hat{\omega} -i\delta} \frac{\Gamma(-2i\delta) 
\Gamma(1-2i\delta) \Gamma(1 -4i\tau/\sigma) }{ (\Gamma(1/2 -2i\hat{\omega} -i\delta))^{2} \Gamma(1/2 +2i\hat{\omega} +i\delta -4i\tau/\sigma) }
 \nonumber \\
&\times&
\left [ { \Gamma(2i\delta) \over \Gamma(-2i\delta)} {\Gamma(1+2i\delta) 
\over \Gamma(1-2i\delta) } \left ( {\Gamma(1/2-2i\hat{\omega}-i\delta) 
\over \Gamma(1/2-2i\hat{\omega}+i\delta) } \right)^2 
+ (-2i\hat{\omega}\sigma)^{2i\delta}{\Gamma(1/2+2i\hat{\omega}+
i\delta-4i\tau/\sigma) \over
\Gamma(1/2+2i\hat{\omega}-i\delta-4i\tau/\sigma) } \right ] \\
A^{\rm out}(\omega) &=& r_{+} (2i\hat{\omega})^{-1/2 +2i\hat{\omega} -i\delta} \sigma^{1/2 -2i\hat{\omega} -i\delta} \frac{ \Gamma(-2i\delta) 
\Gamma(1-2i\delta)}{ \Gamma(1/2 +2i\hat{\omega} -i\delta)
\Gamma(1/2 -2i\hat{\omega} -i\delta) } 
\frac{\Gamma(1-4i\tau/\sigma)}{\Gamma(1/2 +2i\hat{\omega} +i\delta -4i\tau/\sigma)} 
 \nonumber \\
&\times& \left [ \frac{\Gamma(2i\delta) \Gamma(1+2i\delta) \Gamma(1/2 +2i\hat{\omega} -i\delta)
\Gamma(1/2 -2i\hat{\omega} -i\delta)}{ \Gamma(-2i\delta) \Gamma(1-2i\delta) \Gamma(1/2 -2i\hat{\omega}
+i\delta) \Gamma(1/2 +2i\hat{\omega} +i\delta) } +
(2i\hat{\omega}\sigma)^{2i\delta} \frac{ \Gamma(1/2 +2i\hat{\omega} +i\delta -4i\tau/\sigma)}
{\Gamma(1/2 +2i\hat{\omega} -i\delta -4i\tau/\sigma)} \right ]
\label{Ainout}
\end{eqnarray}  
and their ratio will be
\begin{eqnarray}
\frac{A^{\rm out}}{A^{\rm in}} &=& (-1)^{1/2 +2i\hat{\omega} +i\delta} (2i\hat{\omega})^{4i\hat{\omega}} \frac{ \Gamma(1/2 -2i\hat{\omega} -i\delta)}{\Gamma(1/2 +2i\hat{\omega} -i\delta)}
 \nonumber \\
&\times&  \left [ \frac{\Gamma(2i\delta) \Gamma(1+2i\delta) \Gamma(1/2 +2i\hat{\omega} -i\delta)
\Gamma(1/2 -2i\hat{\omega} -i\delta)}{ \Gamma(-2i\delta) \Gamma(1-2i\delta) \Gamma(1/2 -2i\hat{\omega}
+i\delta) \Gamma(1/2 +2i\hat{\omega} +i\delta) } +
(2i\hat{\omega}\sigma)^{2i\delta} \frac{ \Gamma(1/2 +2i\hat{\omega} +i\delta -4i\tau/\sigma)}
{\Gamma(1/2 +2i\hat{\omega} -i\delta -4i\tau/\sigma)} \right ]
 \nonumber \\
 &\times &\left [ { \Gamma(2i\delta) \over \Gamma(-2i\delta)} {\Gamma(1+2i\delta) \over 
\Gamma(1-2i\delta) } \left ( {\Gamma(1/2-2i\hat{\omega}-i\delta) \over
\Gamma(1/2-2i\hat{\omega}+i\delta) } \right)^2 + (-2i\hat{\omega}\sigma)^{2i\delta} 
{\Gamma(1/2+2i\hat{\omega}+i\delta-4i\tau/\sigma) \over 
\Gamma(1/2+2i\hat{\omega}-i\delta-4i\tau/\sigma) } \right ]^{-1}
\label{ratioA1} 
\end{eqnarray} 
In the special case of an extreme Kerr black hole we have $\sigma =0$, 
and  (\ref{ratioA1}) then reduces to
\begin{eqnarray}
\left. \frac{A^{\rm out}}{A^{\rm in}}\right|_{a=M} &=& (-1)^{1/2 +2i\omega M +i\delta} (2i\omega M)^{4i\omega M} \frac{ \Gamma(1/2 -2i\omega M -i\delta)}{\Gamma(1/2 +2i\omega M -i\delta)} \nonumber \\ 
&\times&  \left [ \frac{\Gamma(2i\delta) \Gamma(1+2i\delta) \Gamma(1/2 
+2i\omega M -i\delta)
\Gamma(1/2 -2i\omega M -i\delta)}{ \Gamma(-2i\delta) \Gamma(1-2i\delta) \Gamma(1/2 -2i\omega M
+i\delta) \Gamma(1/2 +2i\omega M +i\delta) } + (8\omega M\tau)^{2i\delta}
 \right ]  \nonumber \\ 
 &\times &\left [ { \Gamma(2i\delta) \over \Gamma(-2i\delta)} {\Gamma(1+2i\delta) \over 
\Gamma(1-2i\delta) } \left ( {\Gamma(1/2-2i\omega M-i\delta) \over
\Gamma(1/2-2i\omega M + i\delta) } \right)^2 + (-8\omega M \tau)^{2i\delta} \right ]^{-1}
\label{ratioAex} 
\end{eqnarray}
Note that this expression is multivalued at $\omega= m\omega_{+}$ due to 
the presence of the $\tau^{2i\delta}$ term (see Appendix B).

 
\section{Analytical properties of solutions of the Teukolsky equation}

In this Appendix we study the analytical properties of the $u^{\rm in}$ 
function in the complex $\omega$-plane. A much more rigorous and 
detailed treatment has been provided  by Hartle and Wilkins \cite{hw}. 
However, as we will show here their 
conclusion that $ u^{\rm in}$ has a branch point for $a< M$ is incorrect.
Instead, we find  that $u^{\rm in}$ has a series of 
(physically insignificant) poles. 
Only for the special case $a=M$ do these poles disappear, 
and $u^{in}$ has indeed a branch point.
 
To begin we write the radial Teukolsky equation (\ref{radeq}) in a 
more compact form (in this Appendix we shall be considering perturbations 
of an arbitrary spin $s$ field)
\begin{equation}
\frac{d^2u}{dr_\ast^2} + Q(r_\ast,\omega) u= 0 \ .
\label{radteuk2}
\end{equation}
It is a well-known fact in scattering theory \cite{newton} that 
certain analytical properties (with relevance for the late-time 
behaviour of the field) of the solutions to an equation of the form 
(\ref{radteuk2}), can be found 
by studying the asymptotic behaviour of these solutions as 
$r_{\ast} \to \pm \infty$.   Therefore, we consider (\ref{radteuk2}) 
in the limit $r \to r_{+}$. For $a<M$ the potential has the asymptotic form, 
\begin{equation}
Q(r_\ast,\omega) \approx (\omega -m\omega_{+} )^2 - V_{\rm h}(\omega)  
+ V_{\rm o}(\omega) e^{cr_{\ast}} \quad  \mbox{as} \quad r_\ast \to -\infty \ ,
\end{equation}
where 
\begin{equation}
V_{\rm h}(\omega)=  2is(\omega -m\omega_{+} )
{(r_{+} - M) \over 2Mr_{+} } + s^2{ (r_{+} - M)^2 \over (2Mr_{+})^2 }  \ ,
\end{equation}
and $ c= (r_{+} -r_{-})/2Mr_{+}$. The explicit form of the  
$V_{\rm o}(\omega)$ function will not be required in the following. 
Setting
\begin{equation}
\Omega= (\omega -m\omega_{+})  - {is(r_{+} -M) \over 2Mr_{+}}  \ ,
\end{equation}
equation (\ref{radteuk2}) becomes 
\begin{equation}
{d^2u \over dr^2_{*}} + \left[ \Omega^2 -
V_{\rm o}e^{cr_{\ast}} \right]u=0 \ ,
\end{equation}
near the horizon.
This equation can be solved by employing the standard Born approximation.
However, it turns out that we can solve it exactly \cite{newton}. 
By setting $ y= e^{cr_{\ast}} $ we can rewrite the equation as
\begin{equation}
y^2 \frac{d^2u}{dy^2} +y\frac{du}{dy} + \frac{1}{c^2} 
\left ( \Omega^2 - V_{\rm o}y \right )u =0
\end{equation}
This is a Bessel-type equation and the solution $u^{\rm in}$ is
\begin{equation}
u^{\rm in}(r_{\ast},\omega)= \left ( \frac{-iV_{\rm o}^{1/2}}{c} \right)^{2i\Omega/c}\Gamma(1- 2i\Omega/c)J_{-2i\Omega/c}
\left( \frac{-2iV_{\rm o}^{1/2}}{c}e^{cr_{\ast}/2} \right)
\label{inB1}
\end{equation}
From the  small argument approximation for the Bessel function
(recall that $r_\ast \to -\infty$ as we approach the horizon), it is 
easy to see that this solution indeed satisfies the appropriate boundary behaviour at the horizon: $u^{\rm in} \to \Delta^{-s/2}e^{-ikr_{\ast}}$.
The presence of the $\Gamma$ function signals the existence of simple poles
at the points $ 1- 2i\Omega/c =-n $ (where $n$ a non-negative integer) 
or, equivalently, at frequencies 
\begin{equation}
\omega= m\omega_{+} + i(s -n-1)\frac{r_{+} -r_{-}}{4Mr_{+}}
\label{falsepoles}
\end{equation}

If we had chosen to use the Born approximation, then at leading order, we 
would had picked up the first of these poles. The existence of these 
poles can be directly deduced by inspection of equation (\ref{smallx}).    
These poles have, however, no physical significance.  The black hole Green's 
function does  not inherit them  since the factor $\Gamma(1- 2i\Omega/c) $ will
be cancelled by an identical factor in the Wronskian (in the $r_{\ast} \to +\infty$ form of $u^{\rm in}$ these poles are contained in the amplitudes $A^{\rm in}$, $A^{\rm out}$, see eqs. (\ref{A}), (\ref{B}) of Appendix A ).
As a consequence, these are called ``false'' poles in quantum scattering theory  \cite{newton}.

Due to the appearance of the Bessel function  of a noninteger power in
(\ref{inB1}), one would expect $u^{\rm in}$ to have branch points at 
those frequencies $\omega_b$ that solve $V_{\rm o}(\omega_b)= 0$. 
However, since
$ J_{\rm \nu}(ze^{2\pi i})= e^{2\pi\nu i} J_{\rm \nu}(z)$ \cite{stegun}
we are effectively left with a single-valued function. It is a quite
general result in scattering theory that exponentially decaying potentials
cannot lead to the existence of branch points \cite{newton}.     
           
We now turn to the $a=M$ case. Approximating the effective radial potential
near the horizon for $a=M$ yields 
\begin{equation}
Q(r,\omega) \approx (\omega -m\omega_{+})^2 -2(\omega-m\omega_{+})
\frac{(m -is)}
{r_\ast}  \quad \mbox{as} \quad r_\ast \to -\infty \ .
\end{equation} 
Near the horizon then, eqn. (\ref{radteuk2}) effectively becomes
\begin{equation}
\frac{d^2u}{dr_{\ast}^2} + \left ( \tilde{\Omega}^2 + \frac{\beta}{r_{\ast}} 
\right ) u =0
\label{whittaker}
\end{equation}
with $\tilde{\Omega}= \omega -m\omega_{+}$ and 
$\beta= -2(m-is)\tilde{\Omega}$.
This equation can be easily transformed to a Whittaker equation \cite{stegun}
with two independent solutions $ W_{\kappa,\mu}(2i\tilde{\Omega}r_{\ast}),
W_{-\kappa,\mu}(-2i\tilde{\Omega}r_{\ast}) $ where 
$\kappa= im +s, ~ \mu=\pm 1/2 $. The Whittaker function $W_{\kappa,\mu}(z) $ 
is defined by,
\begin{eqnarray}
W_{\kappa,\mu}(z) &=& z^{1/2 + \mu} e^{-z/2} 
U(1/2 + \mu -\kappa, 1+ 2\mu, z) \ ,
\\
\label{whit}
\end{eqnarray}
Taking a large argument approximation for the confluent hypergeometric 
functions we deduce that $W_{\kappa,\mu}(2i\tilde{\Omega}r_{\ast})$ is the 
solution with the desired ``ingoing wave'' behaviour at the horizon. 
Clearly, $W_{\kappa,\mu}(z)$ is a multi-valued function
with a branch point at $z=0$, which in the present case corresponds to
$\omega= m\omega_{+}$.
This is in agreement with the result of Hartle and Wilkins \cite{hw}.
In this case the branch point {\em is} inherited by the relevant Green's function.
More specifically, it is contained in the ratio $A^{\rm out}/A^{\rm in},~$
see equation (\ref{ratioAex}).
 
\section{Gravitational perturbations}

So far, we have mainly considered the model problem of a 
massless scalar field. We did this for reasons of simplicity. 
However, it turns out that our results  are easily extended to 
the physically more interesting case of gravitational perturbations.
In fact, the calculation proceeds very much along the same lines.
This means that all conclusions reached in the 
main part of this paper remain valid also for the gravitational case.

In the Teukolsky formalism, gravitational perturbations are represented
by the the scalar functions ${}_{s}\Psi$ with $s= \pm 2$ \cite{teuk1}. 
As before, we can separate the dependence on the $\varphi$ angle 
\begin{equation}
{}_{s}\Psi= \Delta^{-s/2}(r^2+a^2)^{-1/2} \sum_{m=-\infty}^{+\infty}
{}_{s}\psi_{m}(r,\theta,t) e^{im\varphi}
\end{equation}
Further decomposition is possible in the Fourier domain
\begin{equation}
{}_{s}\psi_{m}(r,\theta,t)= {1\over 2\pi} \sum_{l={\rm max}(|m|,|s|)}^\infty
\int_{-\infty+ic}^{+\infty+ic}
{}_{s}R_{l m}(\omega,r) {}_{s}S_{l m}(\theta,a\omega)
e^{-i\omega t} d\omega \ .
\label{gravmfield}
\end{equation}
where ${}_{s}R_{l m}$ solves (\ref{radeq}) with the potential,
\begin{equation}
Q(r,\omega)= {K^2 -2is(r-M)K  + ( 4i\omega r s -2am\omega -a^2\omega^2 - E)\Delta \over (r^2+a^2)^2} 
-{dG \over dr_\ast} -G^2
\end{equation}
with $G= s(r-M)/(r^2 + a^2) + r\Delta/(r^2+a^2)^2$.
The solutions of the homogeneous equation (\ref{radeq}) which are used
for the construction of the Green's function are now
\begin{equation}
{}_{s}u^{\rm up} \sim \left\{ \begin{array}{ll} {}_{s}D^{\rm out}
\Delta^{s/2} e^{ikr_\ast} + {}_{s}D^{\rm in} \Delta^{-s/2} e^{-ikr_\ast}
\quad \mbox{as } r\to r_+ \ , \\ r^{-s}e^{i\omega r_\ast} \quad 
\mbox{as } r\to +\infty \ , \end{array} \right.
\label{gravup}
\end{equation}
and
\begin{equation}
{}_{s}u^{\rm in} \sim \left\{ \begin{array}{ll}
\Delta^{-s/2}e^{-ikr_\ast} \quad \mbox{as } r\to r_+ \ , \\ 
{}_{s}A^{\rm out} r^{-s} e^{i\omega r_\ast} + {}_{s}A^{\rm in} r^s 
e^{-i\omega r_\ast} \quad
\mbox{as } r\to +\infty \ .
\end{array} \right.
\label{gravin}
\end{equation}
Repeating the same steps as in Section IIA, we end up with an expression
similar to (\ref{mfield2})
\begin{equation}
{}_{s}\psi_{m}(r,\theta,t) \approx -\frac{1}{2\pi r^s} \sum_{l= {\rm max}(|m|,|s|)}^{\infty} \int_{-\infty+ic}^{+\infty+ic} \frac{e^{-i\omega(t-r_\ast)}}
{2i\omega}{}_{s}S_{lm}^{a\omega}(\theta) 
\left [ \frac{{}_{s}A^{\rm out}}{{}_{s} A^{\rm in}} 
{}_{s}{\cal I}_{l m}^{+}(\omega)
+ {}_{s}{\cal I}_{l m}^{-}(\omega) \right ]
\label{gravmfield2}
\end{equation}
The analytical properties of the Green's function are identical to the scalar
case (recall that the analysis in Appendix B is valid for an arbitrary spin
field). The long-lived QNMs are now solutions of \cite{detweiler}
\begin{equation}
-\frac{ \Gamma(2i\delta)\Gamma(1+2i\delta)
\Gamma(1/2 +s-2i\hat{\omega}-i\delta)\Gamma(1/2-s-2i\hat{\omega}-i\delta)}
{\Gamma(-2i\delta)\Gamma(1-2i\delta)\Gamma(1/2 +s-2i\hat{\omega}+i\delta)
\Gamma(1/2 -s-2i\hat{\omega}+i\delta)}= (-2i\hat{\omega}\sigma)^{2i\delta} 
{\Gamma(1/2+2i\hat{\omega}+i\delta-4i\tau/\sigma) \over
\Gamma(1/2+2i\hat{\omega}-i\delta-4i\tau/\sigma) } \ .
\label{gravmodes}
\end{equation}
This equation turns out to be invariant under the change $s \to -s$. 
It thus follows  that the  $s=\pm 2$ components share the same set
of long-lived QNMs.

We need to find approximate ($a\approx M, \omega \approx m\omega_{+}$)
expressions for the coefficients ${}_{s}A^{\rm in}, {}_{s}A^{\rm out}$.
Press and Teukolsky \cite{teuk2} give the relevant
approximate radial wavefunctions in a Kerr-ingoing coordinates. 
We need to relate their
solutions to the respective wavefunctions ${}_{s}u^{\rm in},{}_{s}u^{\rm up}$
in order to extract the asymptotic amplitudes
 ${}_{s}A^{\rm in}$ and  ${}_{s}A^{\rm out}$.   
In the standard Boyer-Lindquist coordinates, a spin-$s$ field  can be 
decomposed as
\begin{equation}
{}_{s}\Psi= {}_{s}{\cal R}_{l m}(r,\omega) 
{}_{s}S_{l m}^{a\omega}(\theta)
e^{im\varphi -i\omega t} \ .
\end{equation}
Similarly, in Kerr-ingoing coordinates, we get
\begin{equation}
{}_{s}\tilde{\Psi}= {}_{s}\tilde{{\cal R}}_{l m}(r,\omega) 
{}_{s}S_{l m}^{a\omega}
(\theta) e^{im\tilde{\varphi}} e^{-i\omega \upsilon} \ .
\end{equation}  
These two representations are related by \cite{teuk2}
\begin{equation}
{}_{-s}\Psi= \left ( \frac{\Delta}{2} \right )^{s} {}_{s}\tilde{\Psi} \ .
\label{rel0}
\end{equation}
We adopt the following normalisation for the ``in'' radial wavefunctions,
at $r\to \infty$ (hereafter, for brevity, we drop the $l m$ subscript) 
\begin{equation}
{}_{s}{\cal R}^{\rm in} \sim \left\{ \begin{array}{ll}
\Delta^{-s}e^{-ikr_\ast} \quad \mbox{as } r\to r_+ \ , 
\\ 
\quad {}_{s}B^{\rm in} r^{-1} e^{-i\omega r_{\ast}} + 
{}_{s} B^{\rm out} r^{-2s-1} e^{ i\omega r_{\ast}} \quad
\mbox{as } r\to +\infty \ .
\end{array} \right.
\label{norm}
\end{equation}
\begin{equation}
{}_{s}\tilde{{\cal R}}^{\rm in} \sim \left\{ \begin{array}{ll}
1 \quad \mbox{as } r\to r_+ \ , 
\\ 
\quad {}_{s}Z^{\rm in} r^{-2s-1} + {}_{s}Z^{\rm out} 
r^{-1} e^{2i\omega r_{\ast}}   \quad
\mbox{as } r\to +\infty \ .
\end{array} \right.
\label{innorm}
\end{equation} 
We are mainly interested in the $s=-2$ component, as it leads to the 
dominant contribution to the gravitational waves 
reaching infinity \cite{teuk1}. Inspection of (\ref{gravin})
 and  (\ref{norm}) gives,
\begin{equation}
\frac{ {}_{-2}A^{\rm in} }{{}_{-2}A^{\rm out}} = 
\frac{ {}_{-2}B^{\rm in} }{ {}_{-2}B^{\rm out} }
\label{AnB}
\end{equation}  
From (\ref{rel0}) and (\ref{norm}) we obtain,  
\begin{equation}
{}_{-2}{\cal R}^{\rm in}= f \Delta^{2} {}_{2} \tilde{{\cal R }}^{\rm in}
\exp\left[  -i\int \frac{K}{\Delta} dr \right]
\label{rel1}
\end{equation}
where $f$ some proportionality factor. Furthermore, it is possible to 
relate the two $s=-2$ fields, by means of the Starobinsky identity \cite{teuk2}:
\begin{equation}
{}_{-2}{\cal R}_{lm}(r,\omega)= \frac{\Delta^2}{4|C|^2} J_{+}^{4} 
[ \Delta^2 {}_{2}{\cal R}_{lm}(r,\omega) ]
\label{starob}
\end{equation}
where $ J_{+}= d/dr + iK/\Delta $ and
\begin{equation}
|C|^2 = (\lambda^2 + 4a\omega m -4a^2 \omega^2)[(\lambda-2)^2 + 36a\omega m 
- 36a^2\omega^2 ] +(2\lambda-1)(96a^2 \omega^2 -48a\omega m) + 144\omega^2(M^2-a^2)
\end{equation}
with $\lambda = E + a^2\omega^2 -2a\omega m$.
Combining (\ref{rel1}) and (\ref{starob}) we get
\begin{equation}
{}_{-2}{\cal R}= f \frac{\Delta^2}{4|C|^2}\frac{d^4}{dr^4}
({}_{-2}\tilde{{\cal R}}) e^{-i\int \frac{K}{\Delta} dr}
\label{rel2}
\end{equation}
Eqns. (\ref{rel1}) and (\ref{rel2}) can be used to relate the 
coefficients (${}_{-2}A^{\rm in},{}_{-2}A^{\rm out}$) to (${}_{2}Z^{\rm in},
{}_{2}Z^{\rm out}$) and  (${}_{-2}Z^{\rm in},{}_{-2}Z^{\rm out}$), 
respectively. The first of these connecting formulas simply yields 
${}_{-2}A^{\rm in}/{}_{-2}A^{\rm out} = {}_{2} Z^{\rm in}/{}_{2} Z^{\rm out}$. The QNM frequencies therefore satisfy
\begin{equation}
\left. \frac{ {}_{-2}A^{\rm in} }{{}_{-2}A^{\rm out} }\right|_{\omega=\omega_n}=
\left. \frac{ {}_{2} Z^{\rm in} }{ {}_{2}Z^{\rm out} }\right|_{\omega =\omega_n}= 0
\end{equation}
This equation yields (\ref{gravmodes}) for $s=2$, which is, as pointed out, invariant under the operation $s\to -s$. It is easy to verify that the 
 long-lived gravitational QNMs are given by an expression similar to
(\ref{llmodes}). We have,
\begin{eqnarray}
\frac{{}_{-2}A^{\rm out}}{{}_{-2}A^{\rm in}}(\omega) &=&
r_{+}^{-4} (2i\hat{\omega})^{4+4i\hat{\omega}} \frac{\Gamma(-3/2 -2i\hat{\omega}
-\delta)}{\Gamma(5/2 + 2i\hat{\omega} -i\delta)}
\nonumber 
\\
&\times&
\left [ { \Gamma(2i\delta) \over \Gamma(-2i\delta)} {\Gamma(1+2i\delta) \over \Gamma(1-2i\delta) } 
{\Gamma(5/2-2i\hat{\omega}-i\delta)\Gamma(-3/2-2i\hat{\omega}-i\delta) \over 
\Gamma(5/2-2i\hat{\omega}+i\delta)\Gamma(-3/2-2i\hat{\omega}+i\delta) }  
+ (-2i\hat{\omega}\sigma)^{2i\delta} {\Gamma(1/2+2i\hat{\omega}+i\delta-4i\tau/\sigma) \over
\Gamma(1/2+2i\hat{\omega}-i\delta-4i\tau/\sigma) } \right ]^{-1}
\nonumber 
\\
&\times&  \left [ \frac{\Gamma(2i\delta) \Gamma(1+2i\delta) \Gamma(5/2 +2i\hat{\omega} -i\delta) \Gamma(5/2 -2i\hat{\omega} -i\delta)}
{ \Gamma(-2i\delta) \Gamma(1-2i\delta) \Gamma(5/2 -2i\hat{\omega}
+i\delta) \Gamma(5/2 +2i\hat{\omega} +i\delta) } +
(2i\hat{\omega}\sigma)^{2i\delta} \frac{ \Gamma(1/2 +2i\hat{\omega} +i\delta -4i\tau/\sigma)}
{\Gamma(1/2 +2i\hat{\omega} -i\delta -4i\tau/\sigma)} \right ]
\end{eqnarray}    
It follows that the analysis presented in Section IIE can be 
 used also in the gravitational case, yielding results indentical to the 
scalar ones.

We note that the gravitational perturbation problem can alternatively
be approached exclusively via the $s=-2$ fields, see equation (\ref{rel2}).
This choice was made by Sasaki and Nakamura \cite{sasaki} in their
study of the late-time response of an extreme Kerr hole 
perturbed by an orbiting  test particle. Their results do not agree
with ours in that they find that the late-time signal decays as
$\sim 1/t^{1/2}$. Such a power-law would lead to 
an (unphysical) divergent flux integral \cite{sasaki}. 
We believe that the calculation of Sasaki and Nakamura contains an error in 
the relation between the relevant asymptotic amplitudes. According to their
result, the ratio of the asymptotic amplitudes
$ {}_{-2} B^{\rm in}/{}_{-2}B^{\rm out} $ introduces additional powers of 
$\tau$ in the integrand of (\ref{gravmfield2}) which then leads to 
the different 
power-law behaviour. We have perfomed the calculation (using equations
(\ref{rel2}), (\ref{largex})) and found
\begin{equation}
\frac{ {}_{-2} B^{\rm in} }{ {}_{-2}B^{\rm out} }= r_{+}^{8} D(\omega)
\frac{ {}_{-2} Z^{\rm in} }{ {}_{-2}Z^{\rm out} }
\end{equation}
with
\begin{equation}
D(\omega)= (2i\hat{\omega})^{-4} \left ( (2\hat{\omega} +\delta)^4 +
\frac{5}{2}(2\hat{\omega} +\delta)^2 + \frac{9}{16} \right )
\left ( (2\hat{\omega} - \delta)^4 + \frac{5}{2}(2\hat{\omega} -\delta)^2 + \frac{9}{16} \right )
\end{equation}
which differs crucially from the result of Sasaki and Nakamura.
However, given that several derivatives of the asymptotic solutions are required to 
derive this result, and that significant cancellations occur, it would be very easy to make a mistake
in this calculation. 
  


\pagebreak

\begin{table}
\begin{tabular}{ccccc}
$n$   & \mbox{Re }$\omega_n M$  & \mbox{Im }$ \omega_n M $ & \mbox{Re }$(M A^{\rm out}/\alpha_n)$ 
& \mbox{Im }$(M A^{\rm out}/\alpha_n)$ \\  
\hline
   0 &  0.99324 &  -0.00341 &   0.00216 & -0.00369 \\
   1 &  0.99322 &  -0.01020 &  -0.00691 & -0.00928 \\ 
   2 &  0.99321 &  -0.01699 &  -0.02167 &  0.01501 \\   
   3 &  0.99320 &  -0.02385 &  -0.02167 &  0.01501 \\ 
   4 &  0.99317 &  -0.03067 &  -0.01288 &  0.03089 \\ 
   5 &  0.99313 &  -0.03749 &   0.00398 &  0.04030 \\ 
   6 &  0.99308 &  -0.04432 &   0.02439 &  0.040882 \\ 
   7 &  0.99303 &  -0.05115 &   0.04427 &  0.03253 \\ 
   8 &  0.99298 &  -0.05798 &   0.06043 &  0.01643 \\
   9 &  0.99291 &  -0.06482 &   0.07055 & -0.00550 \\ 
\end{tabular}

\caption{Frequencies and excitation coefficients for the first ten 
$l=m=2$  QNMs, for an $a=0.9999M$ Kerr 
black hole. These results illustrate the fact that many QNMs are 
excited to a comparable amplitude (cf. the excitation coefficients listed
in the last two columns). }
\label{tab1}
\end{table}              

\end{document}